\pdfoutput=1
\documentclass[
preprintnumbers,
preprint,
a4paper,
amsmath, 
amssymb, 
floatfix,
nofootinbib,
]{revtex4}

\pdfoutput=1
\usepackage{graphicx}

\begin{document}

\title{Distorted static photon surfaces in perturbed Reissner-Nordstr\"om spacetimes}

\author{Hirotaka Yoshino${}^{1,2}$}

\affiliation{${}^1$Department of Physics, Osaka Metropolitan University, Osaka 558-8585, Japan}

\affiliation{${}^2$Nambu Yoichiro Institute of Theoretical and Experimental Physics (NITEP),
Osaka Metropolitan University, Osaka 558-8585, Japan}

\preprint{OCU-PHYS-585, AP-GR-194}

\date{September 25, 2023}

%
%
\begin{abstract}

  The photon surface is defined as a timelike surface $S$
  such that any photon emitted in arbitrary tangential direction 
  to $S$ from an arbitrary point on $S$ continues to propagate on $S$. 
  In this paper, we examine whether a static photon surface can be present
  in distorted electrovacuum spacetimes
  with perturbative approach, by 
  constructing analytic solutions
  to the equations for static perturbations 
  of a Reissner-Nordstr\"om spacetime 
  that are regular outside the background photon surface. 
  For each of the $\ell\ge 2$ modes,
  there are two physical solutions to the perturbative equations
  that correspond to adding the multipole moments of mass and electric charge,
  respectively. By adjusting the ratio of the amplitudes of the two solutions
  appropriately, it is possible to realize a distorted photon surface.
  In the case of $\ell=1$, 
  although there is only one physical solution to the perturbative equations,
  there is also a degree of freedom to shift the coordinate
  position of the photon surface, and this enables
  the formation of a photon surface. Such a photon surface
  has a spherically symmetric spatial section, but it is also distorted
  in the three-dimensional sense. 
  Our results show that distorted static photon surfaces
  can be formed 
  at least at the level of the first-order perturbations,
  and imply that the uniqueness of the photon surface
  may not hold in electrovacuum spacetimes.
\end{abstract}


\maketitle

%
%

\section{Introduction}
\label{Sec:Introduction}

Circular orbits around a spherically symmetric black hole
play important roles in black hole physics.
If light sources are located at distant place,
the image of the neighborhood of the black hole
becomes a black hole shadow whose outer boundary 
corresponds to such circular orbits, although in actual shadows observed
by the Event Horizon Telescope collaborations,
the outer boundary of the shadow may correspond to
the inner edge of the accretion disk \cite{Akiyama:2019,EventHorizonTelescope:2019,EventHorizonTelescope:2022}. 
A collection of worldlines of photons in circular orbits
constitutes a timelike hypersurface called a photon
sphere \cite{Virbhadra:1999,Claudel:2000}. 
The concept of the photon sphere is defined for spherically symmetric
static spacetimes. It is natural to consider the generalization
of this concept. One of them is the 
photon surface defined by Claudel {\it et al.}~\cite{Claudel:2000}.
A photon surface is a timelike surface $S$ such that any photon emitted
in any tangential direction to $S$ from any
point on $S$ propagates on $S$.
The photon surface is the generalization of the photon sphere
in the sense that it can be dynamical and need not be spherically symmetric.
The properties of the photon surface have been
extensively studied \cite{Gibbons:2016,Cederbaum:2019,Koga:2019,Koga:2020,Koga:2022}.

An interesting problem is in which situations 
the photon surface can be formed. 
Naive expectation is that since the condition
for the photon surface is fairly strong, 
the situation where the photon surface is present
is limited. Actually, there is an expectation
that static closed regular photon surfaces
can exist only in highly symmetric spacetimes
(but see also \cite{Koga:2019}).
In other words, there might be the uniqueness
for the static photon surface. 
Cederbaum considered the following uniqueness problem \cite{Cederbaum:2014}:
``Suppose that a static regular photon surface be present  
in an asymptotically flat static vacuum spacetime which is regular on
and outside the photon surface. Is such a spacetime restricted to a
Schwarzschild spacetime?'' This problem is equivalent to 
whether a photon surface can be formed in a distorted vacuum spacetime.
Note that in this problem the regularity of the spacetime
is imposed just on and outside a photon surface.
We allow the possibility that there might be a 
curvature singularity or unusual matter distribution 
inside the photon surface.
Up to now, a complete proof for this problem has not been given.
In order to show the uniqueness, Cederbaum imposed one additional boundary
condition on the photon surface, that is, the constancy
of the lapse function (the square root of
the absolute value of the $tt$ component of the metric).
The photon surface on which the lapse function is constant
is redefined as the photon sphere, and the uniqueness theorem of the 
photon sphere was proved.

There are many generalizations of the uniqueness theorem
in various setups \cite{Cederbaum:2015a,Yazadjiev:2015a,Cederbaum:2015b,Yazadjiev:2015b,Yazadjiev:2015c,Rogatko:2016,Tomikawa:2016,Tomikawa:2017,Shinohara:2021,Yazadjiev:2021}. 
Among them, the uniqueness of the (redefined) photon sphere
in asymptotically flat static electrovacuum spacetimes
is closely related to the
present paper. In this context, the photon sphere
is defined as the photon surface on which both the lapse function and
the electrostatic potential are constant.
There are two different theorems proved by different authors.
The theorem proved in Ref.~\cite{Yazadjiev:2015a}
states that an 
electrovacuum spacetime with a non-degenerate photon sphere 
with one component of spherical topology is isometric
to the Reissner-Nordstr\"om spacetime with $Q^2/M^2<9/8$
(this is the parameter range where the photon sphere 
exists in the Reissner-Nordstr\"om spacetimes, see Sec.~\ref{Sec:PS-RN}),
where $Q$ is the electric charge and $M$ is the mass.
The proof of Ref.~\cite{Cederbaum:2015b} allows the possibility that
a photon sphere has multiple components. Under the assumption
of sub-extremality, the spacetime is shown to be isometric
to the Reissner-Nordstr\"om spacetime with $Q^2/M^2< 1$.

As seen above, in the existing uniqueness theorems,
we need to impose the constancy of the lapse function 
in the vacuum case, and the constancy of the lapse function
and the electrostatic potential 
in the electrovacuum case, because of the technical reason.
It is interesting to consider what happens
if we omit these assumptions: we would like to
consider whether the uniqueness holds for a photon
surface in vacuum and electrovacuum spacetimes.
For the vacuum case, the present author has given
a partial answer using the perturbative method \cite{Yoshino:2016}.
In that paper, static perturbations of a Schwarzschild spacetime
were studied, and whether the photon surface can be present
was examined. It was clarified that the photon surface
at $r=3M$ disappears once the spacetime is distorted
if the outside region is vacuum.
This means that there is no regular solution sequence of
regular vacuum spacetimes possessing photon surfaces 
that connects to the Schwarzschild solution. 
In addition, it was also pointed out that if matter is present
in the outside region,
the photon surface could form although fine-tuned distribution
of matter is required.

The perturbative method is very powerful in order to judge whether
some solution sequence exists or not.
For example, by the absence of a regular static perturbation for the 
Schwarzschild-Tangelini spacetime,
a solution sequence of distorted static
vacuum spacetimes that connects to the Schwarzschild-Tangelini
solution has been excluded \cite{Kodama:2004}.
Conversely, if a solution to the static perturbation is found,
there would be 
an unknown static solution sequence,
although its existence finally must be determined
by taking account of nonlinearity of the Einstein equations.
For example, the onset of the
Gregory-Laflamme instability is a static perturbation,
and it indicated the existence of the nonuniform black string solution \cite{Gregory:1993}.
Later, the solutions of the nonuniform black strings were
constructed by fully numerical calculations \cite{Wiseman:2002,Kleihaus:2006}.

The purpose of this paper is to
extend our previous perturbative study
for a Schwarzschild spacetime to the
case of a Reissner-Nordstr\"om spacetime.
We will clarify 
the fact that a photon surface can be present
in a perturbed Reissner-Nordstr\"om spacetime,
in contrast to the Schwarzschild case. 
Specifically, we construct analytic solutions to the
perturbative equations. 
For each of the $\ell\ge 2$ modes, 
there are two independent solutions,
which are interpreted as
the distortion caused by adding the multipole moments of mass and electric charge,
respectively. 
Due to the existence of two independent solutions,
the existence of a photon surface can be realized
by tuning the ratio of the 
two amplitudes. For the $\ell=1$ mode, 
there is only one independent physical solution.
It will be shown that a photon surface, which is distorted
in three-dimensional sense, can exist 
by adjusting its position appropriately. 
Our result indicates that 
the uniqueness of the photon surface does not hold
in electrovacuum spacetimes at least at the level of  
first-order perturbation,
and that the uniqueness may not hold at fully nonlinear level.

This paper is organized as follows.
In the next section, we review the conditions
for a timelike hypersurface $S$ to be a photon surface
and discuss the photon surface in Reissner-Nordstr\"om spacetimes.
In Sec.~\ref{Sec:Static-perturbation-RN}, we consider static perturbations
of a Reissner-Nordstr\"om spacetime 
and derive the perturbative Einstein-Maxwell equations. 
In Secs.~\ref{Sec:Solving_the_perturbation_equations}
and \ref{Sec:Solving_equations_L1}, analytic solutions to
the perturbative equations are derived
for $\ell\ge 2$ and $\ell=1$, respectively. 
In Sec.~\ref{Sec:Distortion-PS}, we show that a distorted photon surface
can be formed in the perturbed Reissner-Nordstr\"om spacetimes.
Section~\ref{Sec:conclusion} is devoted to a summary and discussions.
In Appendix~\ref{Appendix_A},
the formulas for the associated Legendre functions
of the first and second kinds are presented.
We present the specific formulas
for particular solutions $Y_{\rm pt}$ for the perturbative
equations in Appendix~\ref{Appendix_B}.
In Appendix~\ref{Appendix_C},
the perturbative solutions for the electrostatic potential
are presented. 
Throughout the paper, the units in which 
$c = G = 4\pi\varepsilon_0 = 1$ are used, 
where $c$ is the speed of light,
$G$ is the Newtonian constant of gravity, and
$\varepsilon_0$ is the vacuum permittivity.

%
%
\section{Photon surfaces and Reissner-Nordstr\"om spacetimes}
\label{Sec:PS-RN}

In this section, we review the concept of the photon surface
and discuss the photon surfaces in the Reissner-Nordstr\"om spacetimes.

\subsection{Photon surface condition}

As stated in Sec.~\ref{Sec:Introduction},
a photon surface is a timelike surface $S$ such that any photon emitted
in any tangential direction to $S$ from any
point on $S$ propagates on $S$.
In the definition by Claudel {\it et al.} \cite{Claudel:2000},
the photon surface is defined locally.
Therefore, in addition to the photon sphere $r=3M$ in a Schwarzschild
spacetime, the plane $x=\mathrm{constant}$ in a Minkowski spacetime
is also a photon surface. The photon surface can be dynamical,
and the hyperboloid in a Minkowski spacetime is another example of 
the photon surface. 
On the other hand, an event horizon is not a photon surface
because it is a null surface. Also, there is no photon surface
in a Kerr spacetime because although there are null geodesics that
stay on a $r=\mathrm{constant}$ surface, not all null geodesics
emitted tangentially to that surface
propagate on that surface \cite{Teo:2003,Perlick:2004}.

In the context of the uniqueness theorem, we consider a static
photon surface which is topologically closed and completely regular
in an asymptotically flat spacetime.
The known examples of static photon surfaces in this sense are
all spherically symmetric. 
There is an interesting example of a photon surface which
is not spherically symmetric and not completely regular in the spacetime of 
the $C$-metric \cite{Gibbons:2016}. 
Since there is a conical singularity at the pole on this surface,
this surface is a photon surface in the original definition
by Claudel {\it et al.} but not a static photon
surface in the context of the uniqueness theorem.
It was also pointed out in \cite{Koga:2019} that less symmetric
phtoon surfaces are formed in spacetimes that are not asymptotically flat.

We review the conditions for a timelike hypersurface
$S$ in a spacetime $(M,g_{\mu\nu})$ to be a photon surface.
Denoting the unit normal to $S$ as $n^{\mu}$,
the induced metric $h_{\mu\nu}$ is introduced by 
\begin{equation}
h_{\mu\nu}=g_{\mu\nu}-n_{\mu}n_{\nu}.
\end{equation}
The extrinsic curvature $\chi_{\mu\nu}$
on $S$ is defined by $\chi_{\mu\nu} = h_{\mu}^{~\rho}\nabla_{\rho}n_{\nu}$, 
and it can be rewritten as
\begin{equation}
\chi_{\mu\nu} \,=\, \frac12\pounds_nh_{\mu\nu},
\end{equation}
where $\pounds_n$ is the Lie derivative with respect to $n^\mu$.
In Ref.~\cite{Claudel:2000}, three 
necessary and sufficient conditions 
for $S$ to be a photon surface are presented.
The first condition is that any affine-parametrized null geodesic
on the submanifold $S$ is an affine-parametrized null
geodesic in the spacetime $M$ at the same time.
The second condition is that 
\begin{equation}
  \chi_{\mu\nu}k^{\mu}k^{\nu} = 0
  \label{light-surface-condition-2}
\end{equation}
holds for arbitrary null vectors $k^{\mu}$
tangent to $S$.
The third condition is 
\begin{equation}
  \chi_{\mu\nu} \propto h_{\mu\nu},
  \label{light-surface-condition}
\end{equation}
that is, the hypersurface $S$ is umbilical.

Among the three expressions, the umbilical condition
of Eq.~\eqref{light-surface-condition} 
is used in this paper in order 
to examine whether a photon surfaces can be present
in a distorted Reissner-Nordstr\"om spacetime. 
We call the condition of Eq.~\eqref{light-surface-condition}
the ``photon surface condition''.

\subsection{Static photon surfaces in Reissner-Nordstr\"om spacetimes}

The metric of a Reissner-Nordstr\"om spacetime
with mass $M$ and charge $Q$ is
\begin{equation}
d\hat{s}^2=-e^{2\nu^{(0)}}dt^2+e^{2\mu^{(0)}}dr^2+r^2
\left(d\theta^2+\sin^2\theta d\phi^2\right),
\label{SAD1}
\end{equation}
with 
\begin{equation}
e^{2\nu^{(0)}}=e^{-2\mu^{(0)}}= 1 - \frac{2M}{r}+\frac{Q^2}{r^2},
\label{SAD2}
\end{equation}
in the $(t, r, \theta, \phi)$ coordinates.
The vector potential $A_\mu$ of the electromagnetic field
is given by
\begin{equation}
A_\mu \ = \ (-\Phi^{(0)},\, 0, \, 0, \, 0),
\end{equation}
with the electrostatic potential,
\begin{equation}
\Phi^{(0)} \, = \, \frac{Q}{r}. 
\end{equation}
Throughout the paper, we consider the case of the positive mass, $M>0$.
The positivity of the electric charge $Q$ can be assumed
without loss of generality.

%
\begin{figure}
  \centering
  \includegraphics[width=0.5\textwidth,bb= 0 0 260 166]{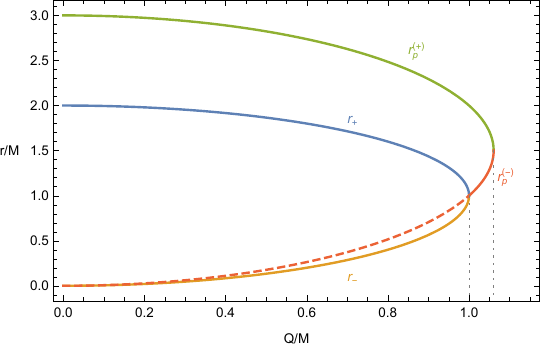}
  \caption{The radius of the event and Cauchy horizons, $r_+$ and $r_-$,
    and that of the outer and inner photon surface, $r_{\rm p}^{(+)}$ and $r_{\rm p}^{(-)}$
    as functions of $Q/M$ in the Reissner-Nordstr\"om spacetime.
    The left and right vertical dotted lines indicate
  $Q/M=1$ and $\sqrt{9/8}$, respectively.}
  \label{Fig:Location-PS}
\end{figure}
%

In the sub-extremal case $M^2>Q^2$, there are the event and Cauchy horizons
at $r=r_+$ and $r_-$, respectively, where
\begin{equation}
r_\pm \, =\, M(1\pm \gamma) 
\end{equation}
with
\begin{equation}
  \gamma \ := \ \frac{\sqrt{|M^2-Q^2|}}{M}.
  \label{Def:Gamma}
\end{equation}
In the extremal case $M^2=Q^2$, there is only one degenerate horizon
at $r=M$. 
In the super-extremal case $M^2<Q^2$, there is no horizon because
$r^2e^{2\nu^{(0)}}$ is strictly positive for $r>0$,
\begin{equation}
r^2e^{2\nu^{(0)}} \, =\, (r-M)^2+M^2\gamma^2.
\end{equation}
In this case, there is a naked curvature singularity at $r=0$
whose characteristic is timelike.
Figure~\ref{Fig:Location-PS} shows the values of
$r_\pm$ as functions of $Q/M$. 

Let us calculate the locations of static photon surfaces
in this spacetime. Let us consider an $r=\mathrm{constant}$
surface. Then, the photon surface condition 
of Eq.~\eqref{light-surface-condition} is equivalent to
\begin{equation}
\frac{\left(e^{2\nu^{(0)}}\right)_{,r}}{e^{2\nu^{(0)}}}\, = \, \frac{2}{r}.
\end{equation}
Solving this equation, we obtain the two solutions,
\begin{equation}
  r\, =\, r_{\rm p}^{(\pm)}\, := \,
  \frac{3M}{2}\left(1\pm \sqrt{1-\frac{8Q^2}{9M^2}}\right).
\end{equation}
These solutions exist only for the situation $Q^2/M^2\le 9/8$.
In the sub-extremal and extremal cases,
the surface $r=r_{\rm p}^{(+)}$ exists outside the event horizon
and it is an observable photon surface.
The surface $r=r_{\rm p}^{(-)}$ is located between the event and Cauchy horizons
in the sub-extremal case, and it is a spacelike umbilical surface
(and hence, not a photon surface).
Since the surface $r=r_{\rm p}^{(-)}$ coincides with the event
horizon in the extremal case, it is not a photon surface.
In the super-extremal case,
there are two observable photon surfaces $r=r_{\rm p}^{(\pm)}$.
The values of $r_{\rm p}^{(\pm)}$ are shown as functions
of $Q/M$ in Fig.~\ref{Fig:Location-PS}.
The above results can be derived by the method of
the effective potential for null geodesics as well.
From the behavior of the effective potential,
the outer photon surface is unstable, and the inner photon surface
is stable against perturbation to the circularly orbiting null geodesics.

Below, we only consider the region around and outside the
outer photon surface in the sub-extremal and extremal cases,
and the region around and  outside the inner photon
surface in the super-extremal case. 
In all cases, the inside region (in particular, in the neighborhood
of the event horizon) is not considered.
To simplify the notation, we denote the 
radial position of the photon surface
(or one of the photon surfaces)
as $r_{\rm p}$, namely, 
\begin{equation}
  r_{\rm p} \, :=\,
  \begin{cases}
    r_{\rm p}^{(+)} &  (0\le Q^2/M^2\le 1),\\
    r_{\rm p}^{(-)}~\textrm{or}~r_{\rm p}^{(+)} &  (1< Q^2/M^2\le 9/8).\\
\end{cases}
\end{equation}
We consider static perturbations which are regular
in the region $r\gtrsim r_{\rm p}$, and 
then, examine whether a photon surface can remain to exist
in continuous sequences of distorted configurations.
In the super-extremal case, both the inner and outer photon
surfaces are studied.

%
%
\section{Static perturbation of a Reissner-Nordstr\"om spacetime}
\label{Sec:Static-perturbation-RN}

In this section, we derive the equations for static perturbations
of a Reissner-Nordstr\"om spacetime.
We consider the spacetime with a metric $g_{\mu\nu}$
and the vector potential $A_{\mu}$. 
Using
the traceless property of the energy-momentum tensor
of the electromagnetic fields, 
the Einstein equation for the electrovacuum spacetime is
\begin{equation}
  R_{\mu\nu} \, = \,
  2F_{\mu\rho}{F_{\nu}}^{\rho}-\frac12g_{\mu\nu}F_{\rho\sigma}F^{\rho\sigma},
  \label{Eq:Einstein-equation}
\end{equation}
where $R_{\mu\nu}$ is the Ricci tensor and $F_{\mu\nu}$
is the electromagnetic field strength given by 
\begin{equation}
F_{\mu\nu} = \nabla_{\mu}A_{\nu}-\nabla_{\nu}A_{\mu},
\end{equation}
with the covariant derivative $\nabla_\mu$.
The Maxwell equation is
\begin{equation}
  \nabla_{\mu}F^{\mu\nu} \, = \, 0,
        \label{Eq:Maxwell}
\end{equation}
where the four-electric current is assumed to be absent
since we consider the electrovacuum spacetime.

\subsection{Static perturbation}

The perturbation of a Reissner-Nordstr\"om spacetime
was formulated by Zerilli \cite{Zerilli:1974} (see also \cite{Kodama:2003})
by extending the Regge-Wheeler-Zerilli formalism for the
perturbation of a Schwarzschild spacetime \cite{Regge:1957,Zerilli:1970}.
In the formulation by Zerilli,
the metric perturbation is decomposed
into the even- and odd-parity perturbations,
and each of them is expanded by the ``tensor spherical harmonics''
on the unit two-sphere 
that are labeled by the angular quantum numbers $(\ell, m)$.
There is also the perturbation of the  
electromagnetic field strength which is also expanded by the tensor spherical harmonics.
In this paper, we consider the perturbation of
the vector potential $A^\mu$ and its expansion
in terms of the spherical harmonics.

We consider static perturbations of a Reissner-Nordstr\"om spacetime.
Such perturbation is 
time-independent even-parity perturbations.
Since we consider the static spacetime,
the $(tr)$-component of the metric is set to be zero.
After a suitable gauge transformation,
the metric can be written in the diagonal form,
\begin{equation} 
d\hat{s}^2=-e^{2\nu}dt^2+e^{2\mu}dr^2+e^{2\psi}r^2
\left(d\theta^2+\sin^2\theta d\phi^2\right), 
\label{metric_deformed}
\end{equation}
with
\begin{subequations}
\begin{eqnarray}
\nu &=& \nu^{(0)}+\epsilon \nu^{(1)}+\cdots,
\label{expand_nu}\\
\mu &=& \mu^{(0)}+\epsilon \mu^{(1)}+\cdots,
\label{expand_mu}\\
\psi &=& \epsilon \psi^{(1)}+\cdots,
\label{expand_psi}
\end{eqnarray}
\end{subequations}
where $\epsilon$ is a small expansion parameter. 
This gauge is called the Regge-Wheeler gauge.
Here, $\nu^{(1)}$, $\mu^{(1)}$, and $\psi^{(1)}$ 
depend only on spatial coordinates. 
The vector potential $A_\mu$ is assumed to
have the form
\begin{equation} 
A_{\mu} = (-\Phi, \, 0, \, 0,\, 0),
\end{equation}
where $\Phi$ is the electrostatic potential that is expanded as
\begin{equation}
\Phi \, =\, \Phi^{(0)} + \epsilon\Phi^{(1)} + \cdots. 
\end{equation}
Here, $\Phi^{(1)}$ is assumed to be independent of $t$.
The fact that $A_\mu$ only has the time component means
that there is a static electric field for static observers.

In the Regge-Wheeler gauge, the first-order functions
are expanded with the spherical harmonics $Y_{\ell}^{m}$ as,
\begin{subequations}
\begin{align}
&\nu^{(1)}=-\sum_{\ell, m}H_{\ell m}^{(1)}(r) Y_{\ell}^{m}(\theta,\phi), 
\label{function_nu1}\\
&\mu^{(1)}=\sum_{\ell, m}L_{\ell m}^{(1)}(r) Y_{\ell}^{m}(\theta,\phi), 
\label{function_mu1}\\
&\psi^{(1)}= \sum_{\ell,m}K_{\ell m}^{(1)}(r) Y_{\ell}^{m}(\theta,\phi),
\label{function_psi1}
\end{align}
\end{subequations}
where $\ell$ and $m$ are integers satisfying
$\ell\ge 0$ and $-\ell\le m\le \ell$.
Similarly, $\Phi^{(1)}$ is expanded as
\begin{equation}
\Phi^{(1)}=\sum_{\ell, m}\varphi_{\ell m}^{(1)}(r) Y_{\ell}^{m}(\theta,\phi).
\end{equation}
Since the first-order equations with different $(\ell, m)$ values
decouple, each mode can be treated separately.
In what follows, we consider a single mode 
and write the radial functions as $H^{(1)}$, $L^{(1)}$, $K^{(1)}$
and $\varphi^{(1)}$ for simplicity.

Similarly to the Birkoff theorem in the vacuum case,
there is the theorem which
states that a spherically symmetric electrovacuum spacetime
must be static \cite{Hoffman}.
In the asymptotically flat case, the solution for the
spherically symmetric static spacetime to the Einstein-Maxwell
equations is limited to the Reissner-Nordstr\"om spacetime. 
Therefore, the $\ell=0$ mode (i.e., spherically symmetric perturbation)
just corresponds to shift in the mass $M$ and the charge $Q$.
For this reason, we do not consider the $\ell=0$ mode.
On the other hand, the $\ell=1$ mode becomes physical
in contrast to the Schwarzschild case,
since it corresponds to adding electric dipole moment to the 
background spacetime.
Therefore, the modes $\ell\ge 1$
are studied below.

As discussed in \cite{Chandrasekhar}, 
due to the spherical symmetry of the background spacetime,
the solutions for nonaxisymmetric modes
can be obtained by rotating the solutions for the axisymmetric modes. 
For this reason, it is sufficient to consider
the axisymmetric perturbation.
In the axisymmetric case, 
the difference 
between the $(\theta\theta)$ and $(\phi\phi)$ components
of the Einstein equation \eqref{Eq:Einstein-equation}
gives
\begin{equation}
  (\mu^{(1)}+\nu^{(1)})_{,\theta\theta} + \cot\theta(\mu^{(1)}+\nu^{(1)})_{,\theta}
  \ = \ 0.
\end{equation}
Integrating this, we find
\begin{equation}
\mu^{(1)}+\nu^{(1)}\, = \, A(r)\cos\theta +B(r).
\end{equation}
This means that $H^{(1)}-L^{(1)}$ only has the modes $\ell=0$ and $1$.
Therefore, for the modes $\ell\ge 2$,
the Einstein equation implies $H^{(1)}=L^{(1)}$.
For the mode $\ell = 1$, $H^{(1)}$ and $L^{(1)}$
are different in general. For this reason,
we have to treat the modes $\ell=1$ and $\ell\ge 2$ separately.
Below, we derive the equations for the
modes $\ell\ge 2$ and $\ell=1$, one by one.

\subsection{Equations for $\ell\ge 2$}

We present the perturbative Einstein-Maxwell equations
for the case $\ell\ge 2$. Using the relation $H^{(1)}=L^{(1)}$, 
the Einstein equations for the first-order quantities are
\begin{subequations}
\begin{equation}
 r^2e^{2\nu^{(0)}}H^{(1)}_{,rr}
+ 2r\left(re^{2\nu^{(0)}}\right)_{,r} H^{(1)}_{,r}
- r^2\left(e^{2\nu^{(0)}}\right)_{,r}K^{(1)}_{,r} 
- \left(\ell^2+\ell-\frac{2Q^2}{r^2}\right) H^{(1)} = 2Q\varphi^{(1)}_{,r},
\label{Rtt}
\end{equation}
\begin{multline}
 r^2e^{2\nu^{(0)}}\left(H^{(1)}_{,rr}-2K^{(1)}_{,rr}\right)
+ 2r\left( re^{2\nu^{(0)}}\right)_{,r} H^{(1)}_{,r} \\
- r\left[ r\left(e^{2\nu^{(0)}}\right)_{,r}+4e^{2\nu^{(0)}}\right] K^{(1)}_{,r}
+ \left(\ell^2+\ell+\frac{2Q^2}{r^2}\right) H^{(1)} = 2Q\varphi^{(1)}_{,r},
\label{Rrr}
\end{multline}
\begin{eqnarray}
e^{2\nu^{(0)}}\left(H^{(1)}_{,r} - K^{(1)}_{,r}\right)
 + \left(e^{2\nu^{(0)}}\right)_{,r}H^{(1)} = \frac{2Q}{r^2}\varphi^{(1)},   
\label{Rr theta}
\end{eqnarray}
\begin{multline}
 r^2e^{2\nu^{(0)}} K^{(1)}_{,rr}
- 2re^{2\nu^{(0)}}H^{(1)}_{,r}
+ r\left[ r\left(e^{2\nu^{(0)}}\right)_{,r}+4e^{2\nu^{(0)}}\right] K^{(1)}_{,r} \\
- 2\left( re^{2\nu^{(0)}} \right)_{,r} H^{(1)} 
-  (\ell^2+ \ell  -2)K^{(1)} = 2Q\varphi^{(1)}_{,r},
\label{R phi phi}
\end{multline}
\end{subequations}
where ${,r}$ denotes the derivative with respect to $r$. 
For $m=0$, these equations are derived from $tt$, $rr$, $r\theta$ components and
the sum of the $\theta\theta$ and $\phi\phi$ components
of the Einstein equations, respectively. 
The Maxwell equation is
\begin{equation}
  r^2e^{2\nu^{(0)}} \left(\varphi^{(1)}_{,rr}+\frac{2}{r}\varphi^{(1)}_{,r}\right)
  -\ell(\ell+1)\varphi^{(1)} =
  \frac{2Q}{r^2}K^{(1)}_{,r}.
  \label{Eq:Maxwell-Lge2}
\end{equation}
The solutions for these equations will be
presented in Sec.~\ref{Sec:Solving_the_perturbation_equations}.

Although there are five equations for the three functions
$H^{(1)}$, $K^{(1)}$, and $\varphi^{(1)}$,
these equations are not overdetermining because
there are two constraint equations, i.e. 
one Hamiltonian constraint and one momentum constraint,
if the Einstein equations are regarded as evolution equations
in the $r$ direction.
In the case of $\ell\ge 2$, the Regge-Wheeler gauge completely
fixes the gauge, and all solutions to these equations
represent physical perturbations (not gauge transformation).

\subsection{Equations for $\ell= 1$}
\label{Sec:Equation-for-L1}

In the case of $\ell = 1$, the condition $H^{(1)}=L^{(1)}$ is
not satisfied in general. The Einstein equations are
\begin{subequations}
\begin{multline}
 r^2e^{2\nu^{(0)}}H^{(1)}_{,rr}
 + \left[\frac32r^2\left(e^{2\nu^{(0)}}\right)_{,r}+2re^{2\nu^{(0)}}\right] H^{(1)}_{,r}
 + \frac12r^2\left(e^{2\nu^{(0)}}\right)_{,r} L^{(1)}_{,r}
 \\
- r^2\left(e^{2\nu^{(0)}}\right)_{,r}K^{(1)}_{,r} 
- 2\left(1-\frac{Q^2}{r^2}\right) H^{(1)} \,=\, 2Q\varphi^{(1)}_{,r},
\label{Rtt-Leq1}
\end{multline}
\begin{multline}
 r^2e^{2\nu^{(0)}}\left(H^{(1)}_{,rr}-2K^{(1)}_{,rr}\right)
+ \frac32r^2\left(e^{2\nu^{(0)}}\right)_{,r} H^{(1)}_{,r} 
+ \left[\frac12r^2\left(e^{2\nu^{(0)}}\right)_{,r}+2re^{2\nu^{(0)}} \right]L^{(1)}_{,r} \\
- r\left[ r\left(e^{2\nu^{(0)}}\right)_{,r}+4e^{2\nu^{(0)}}\right] K^{(1)}_{,r}
+ 2 L^{(1)} + \frac{2Q^2}{r^2}H^{(1)} \,=\, 2Q\varphi^{(1)}_{,r},
\label{Rrr-Leq1}
\end{multline}
\begin{multline}
e^{2\nu^{(0)}}\left(H^{(1)}_{,r} - K^{(1)}_{,r}\right)
+ \left[\frac12\left(e^{2\nu^{(0)}}\right)_{,r}-\frac{e^{2\nu^{(0)}}}{r}\right]H^{(1)}
\\
+ \left[\frac12\left(e^{2\nu^{(0)}}\right)_{,r}+\frac{e^{2\nu^{(0)}}}{r}\right]L^{(1)}
\,=\, \frac{2Q}{r^2}\varphi^{(1)},   
\label{Rrtheta-Leq1}
\end{multline}
\begin{multline}
 r^2e^{2\nu^{(0)}} K^{(1)}_{,rr}
- re^{2\nu^{(0)}}H^{(1)}_{,r}
- re^{2\nu^{(0)}}L^{(1)}_{,r}
+ r\left[ r\left(e^{2\nu^{(0)}}\right)_{,r}+4e^{2\nu^{(0)}}\right] K^{(1)}_{,r} \\
+ \left( 1+\frac{2Q^2}{r^2} \right) H^{(1)} 
-3L^{(1)} 
\,=\, 2Q\varphi^{(1)}_{,r},
\label{Rphiphi-Leq1}
\end{multline}
\end{subequations}
and the Maxwell equation is
\begin{equation}
  r^2e^{2\nu^{(0)}} \left(\varphi^{(1)}_{,rr}+\frac{2}{r}\varphi^{(1)}_{,r}\right)
  -2\varphi^{(1)} \,=\,
  {Q}\left(2K^{(1)}_{,r}+H^{(1)}_{,r}-L^{(1)}_{,r}\right).
  \label{Eq:Maxwell-Leq1}
\end{equation}

Here, we have to comment on the fact that the solutions 
to these equations includes unphysical gauge mode.
If we consider the gauge mode generated by the transformation,
\begin{subequations}
\begin{eqnarray}
r &\to&  r+\epsilon r^2e^{2\nu^{(0)}}b_{,r}\, \sqrt{\frac{3}{4\pi}}\cos\theta,\\
\theta &\to&  \theta+\epsilon b\, \sqrt{\frac{3}{4\pi}}\sin\theta,
\end{eqnarray}
\end{subequations}
where $b=b(r)$ is an arbitrary non-divergent function,
we have
\begin{subequations}
  \begin{eqnarray}
  H^{(1)} &=& -\frac12r^2\left(e^{2\nu^{(0)}}\right)_{,r}b_{,r},\label{gauge-mode-solution-H1}\\  
  L^{(1)} &=& r^2e^{2\nu^{(0)}}b_{,rr}
  +\frac12r\left[r\left(e^{2\nu^{(0)}}\right)_{,r}+4e^{2\nu^{(0)}}\right]b_{,r},\label{gauge-mode-solution-L1}\\  
  K^{(1)} &=& re^{2\nu^{(0)}}b_{,r}+b,\label{gauge-mode-solution-K1}
  \end{eqnarray}
  and
  \begin{equation}
    \varphi^{(1)} \ =\ -Qe^{2\nu^{(0)}}b_{,r}.\label{gauge-mode-solution-phi1}
  \end{equation}
\end{subequations}
Substituting into Eqs.~\eqref{Rtt-Leq1}--\eqref{Rphiphi-Leq1}
and Eq.~\eqref{Eq:Maxwell-Leq1},
it can be checked that these formulas actually satisfy the perturbative
Einstein-Maxwell equations. Using the arbitrariness of $b(r)$,
it is possible to impose the gauge condition
\begin{equation}
  H^{(1)} \ = \ L^{(1)}.
  \label{Gauge-condition-L1}
\end{equation}
In this gauge, Eqs.~\eqref{Rtt-Leq1}--\eqref{Rphiphi-Leq1}
and Eq.~\eqref{Eq:Maxwell-Leq1}
are reduced to Eqs.~\eqref{Rtt}--\eqref{R phi phi}
and Eq.~\eqref{Eq:Maxwell-Lge2} with $\ell=1$.
However, there is still a residual gauge degree of freedom.
Namely, if we choose $b$ that satisfies
\begin{equation}
  \left(r^2e^{2\nu^{(0)}}b_{,r}\right)_{,r} \ = \ 0,
  \label{equation-for-b}
\end{equation}
the perturbation by the gauge transformation, Eqs.~\eqref{gauge-mode-solution-H1} and \eqref{gauge-mode-solution-L1}, 
satisfies Eq.~\eqref{Gauge-condition-L1}.
Integrating Eq.~\eqref{equation-for-b},
we obtain
\begin{equation}
  b=
  \begin{cases}
    \displaystyle \alpha + \frac{\beta}{\gamma}\,\mathrm{arccoth}\left(\frac{r-M}{\gamma M}\right), & (M^2>Q^2);\\
    \displaystyle \alpha + \beta \frac{M}{r-M}, & (M^2=Q^2);\\
    \displaystyle \alpha + \frac{\beta}{\gamma} \,\mathrm{arccot} \left(\frac{r-M}{\gamma M}\right), & (M^2<Q^2),
  \end{cases}
  \label{gauge-mode-function-b}
  \end{equation}
where $\gamma$ is defined in Eq.~\eqref{Def:Gamma}.

To summarize, in the case of $\ell=1$, it is possible to
impose the gauge condition of Eq.~\eqref{Gauge-condition-L1},
and the same equations as the case $\ell\ge 2$,
Eqs.~\eqref{Rtt}--\eqref{R phi phi}
and Eq.~\eqref{Eq:Maxwell-Lge2},
are to be solved. The solutions to these equations
include the gauge mode, and we have to adopt the physical mode
by checking that the solution is not proportional to
the gauge-mode solution discussed above.
In Sec.~\ref{Sec:Solving_equations_L1},
we will obtain the physical solution
to the case $\ell=1$.

%
%
\section{Solving the perturbative equations for $\ell\ge 2$}
\label{Sec:Solving_the_perturbation_equations}

We now solve the perturbative equations
and obtain analytic solutions.
As discussed in the previous case, we are to solve 
Eqs.~\eqref{Rtt}--\eqref{R phi phi}
and Eq.~\eqref{Eq:Maxwell-Lge2}.
In solving these equations, the method of Ref.~\cite{Bini:2006}
is very helpful. We introduce
\begin{subequations}
\begin{eqnarray}
X &:= & H^{(1)} - K^{(1)},\label{Def:X}\\
Y &:= & H^{(1)} + K^{(1)}.\label{Def:Y}
\end{eqnarray}
\end{subequations}
Eliminating $\varphi^{(1)}_{,r}$ from Eqs.~\eqref{Rtt}
and \eqref{R phi phi}, we find
\begin{equation}
  r^2e^{2\nu^{(0)}} X_{,rr}+4(r-M)X_{,r}-(\ell^2+\ell-2)X \,=\, 0.
  \label{Equation-for-X}
\end{equation}
Therefore, the quantity $X$ is determined by a single equation.
Once the solution of $X$ is obtained, the quantity
$Y$ can be determined using the following equation:
\begin{equation}
  r^2e^{2\nu^{(0)}} \left(Y_{,rr}+\frac{2}{r}Y_{,r}\right)
  -\ell(\ell+1)Y
  \, = \,
  r^2e^{2\nu^{(0)}} \left(X_{,rr}+\frac{2}{r}X_{,r}\right)
  +\ell(\ell+1)X,
  \label{Equation-for-Y}
\end{equation}
which is obtained by 
eliminating $\varphi^{(1)}_{,r}$ from Eqs.~\eqref{Rtt}
and \eqref{Rrr}. 
Then, we have the functions $H^{(1)}$ and $K^{(1)}$,
and the quantity $\varphi^{(1)}$ is determined
by Eq.~\eqref{Rr theta}.

Here, we discuss the general strategy
for obtaining the solutions for the cases $\ell\ge 2$.
There are two independent solutions, $X=X_1$ and $X_2$ 
for Eq.~\eqref{Equation-for-X} that behave as
$X_1\sim r^{\ell-1}$ and $X_2\sim 1/r^{\ell+2}$ for large $r$.
The general solution is given by $X=A_1X_1+A_2X_2$.
Since we would like the regular solution outside the
photon sphere, we require $A_1=0$, and hence,
\begin{equation}
X(r)\, = \, A_2\,X_2(r).
\end{equation}
Substituting this solution into Eq.~\eqref{Equation-for-Y},
this equation is symbolically written as
\begin{equation}
  r^2e^{2\nu^{(0)}} \left(Y_{,rr}+\frac{2}{r}Y_{,r}\right)
  -\ell(\ell+1)Y
  \, = \,
  A_2S(X_2).
  \label{Equation-for-Y-rewritten}
\end{equation}
The solution to this equation is the sum of the
homogeneous solutions and the particular solution.
There are two independent homogeneous solutions,
$Y_1$ and $Y_2$ that behave as $Y_1\sim r^{\ell}$
and $Y_2\sim 1/r^{\ell+1}$ at the distant place.
Then, the general homogeneous solution
is written as $Y_{\rm hom}=B_1Y_1+B_2Y_2$.
Since the regularity is imposed outside the photon surface,
$B_1=0$ is required, and hence, we have $Y_{\rm hom}=B_2Y_2$.
Since the source term is proportional to $A_2$,
the particular solution has the form $A_2Y_{\rm pt}$.
Here, there is a degree of freedom to add homogeneous solutions
to $Y_{\rm pt}$, and 
in order to guarantee the regularity, we require
$Y_{\rm pt}$ to decay (at least) as $Y_{\rm pt}\sim 1/r^{\ell+1}$
at the distant place.\footnote{Since there is a degree of freedom
  of adding the homogeneous solution, $Y_{\rm pt}$
  behaves as $Y_{\rm pt}\sim 1/r^{\ell+1}$ in general.
  By adding the homogeneous solution appropriately,
  it is possible to choose a particular solution that behaves
as $Y_{\rm pt}\sim 1/r^{\ell+2}$.} 
To summarize, the solution of $Y$ has the form
\begin{equation}
  Y(r) \, = \, B_2\,Y_2(r) + A_2\,Y_{\rm pt}(r).
 \label{formal-solution-Y}
\end{equation}
Then, from Eqs.~\eqref{Def:X}
and \eqref{Def:Y}, we can calculate $H^{(1)}$ and $K^{(1)}$ as
\begin{subequations}
\begin{eqnarray}
  H^{(1)} &=& A_2\frac{Y_{\rm pt}+X_2}{2} + B_2\frac{Y_2}{2},
  \label{calculate-H1-from-X-Y}\\
K^{(1)} &=& A_2\frac{Y_{\rm pt}-X_2}{2} + B_2\frac{Y_2}{2},
  \label{calculate-K1-from-X-Y}
\end{eqnarray}
\end{subequations}
and the function $\varphi^{(1)}$ can be calculated with Eq.~\eqref{Rr theta}.
It can be checked that $H^{(1)}$, $K^{(1)}$, and $\varphi^{(1)}$
satisfy all perturbative Einstein-Maxwell equations.

Here, we discuss the difference between the perturbation of the
Reissner-Nordstr\"om spacetime 
and that of the Schwarzschild spacetime.
In the case of the Reissner-Nordstr\"om spacetime, 
there are two independent regular solutions in the region $r\gtrsim r_{\rm p}$
whose amplitudes are given by $A_2$ and $B_2$.
Physically, these two modes would correspond to
the distortion induced by adding
the multipole moments of mass and electric charge, respectively.
By contrast, in the Schwarzschild spacetime, 
there is only one physical solution \cite{Yoshino:2016}.
The origin of this difference is that
Eq.~\eqref{Rr theta} is used to determine
$\varphi^{(1)}$ in the Reissner-Nordstr\"om case,
while it gives a constraint to $H^{(1)}$ and $K^{(1)}$,
and thus, kills one of the two solutions in the Schwarzschild case. 
This is the reason why we will arrive at a 
different conclusion from our previous paper \cite{Yoshino:2016}.

Next, we solve for the function $X_2$, $Y_2$, and $Y_{\rm pt}$,
for the extremal, sub-extremal, and super-extremal cases, one by one.

\subsection{Extremal case}
\label{Sec:Extremal-Lge2}

We begin with the extremal case, $M=Q$,
because the analysis is technically simple.
We introduce the rescaled radial coordinate 
with
\begin{equation}
  x=\frac{r}{M}.
  \label{Def:x}
\end{equation}
Equation~\eqref{Equation-for-X} becomes
\begin{equation}
(x-1)^2X_{,xx}+4(x-1)X_{,x}-(\ell^2+\ell-2)X\ = \ 0,
\end{equation}
and the two independent solutions are easily found as
$X_1=(x-1)^{\ell-1}$ and
\begin{equation}
  X_2=\frac{1}{(x-1)^{\ell+2}}.
  \label{X2-extremal-Lge2}
\end{equation}
Then, $X={A_2}X_2$
is adopted for the solution of $X$.
Equation~\eqref{Equation-for-Y-rewritten} becomes
\begin{equation}
  (x-1)^2 \left(Y_{,xx}+\frac{2}{x}Y_{,x}\right)
  -\ell(\ell+1)Y
  \, = \,
  A_2S(X_2),
\end{equation}
with
\begin{equation}
S(X_2)\,=\,  \frac{2}{(x-1)^{\ell+2}}\left[(\ell+1)^2+\frac{\ell+2}{x}\right].
\end{equation}
In order to solve this equation, it is convenient to
introduce $u(x)$ with $Y=[(x-1)/x]u$. Then, the equation is rewritten as
\begin{equation}
  (x-1)^2 u_{,xx}+2(x-1)u_{,x}
  -\ell(\ell+1)u
  \, = \,
  A_2\frac{x}{x-1}S(X_2).
\end{equation}
The two homogeneous solutions are easily found as $u_1=(x-1)^{\ell}$
and $u_2=1/(x-1)^{\ell+1}$, and hence,
$Y_1=(x-1)^{\ell+1}/x$ and
\begin{equation}
  Y_2=\frac{1}{x(x-1)^{\ell}}.
  \label{Y2-extremal-Lge2}
\end{equation}
Then, $B_2Y_2$ is adopted as the homogeneous solution.
The particular solution is calculated by the standard method
(e.g., the Green's function method or the method of variation of constants).
The result is
\begin{equation}
  Y_{\rm pt} \, = \,
  \frac{\ell+1}{x(x-1)^{\ell+1}}
  +\frac{\ell^2+3\ell+3}{(2\ell+3)x(x-1)^{\ell+2}},
  \label{Ypt-extremal-Lge2}
\end{equation}
and the solution of $Y$ in the form of Eq.~\eqref{formal-solution-Y}
is obtained.

The obtained solution is written as
\begin{subequations}
  \begin{eqnarray}
    H^{(1)} &=& A_2\, \frac{(\ell+2)[(2\ell+3)x-\ell]}{2(2\ell+3)x(x-1)^{\ell+2}} + \frac{B_2}{2x(x-1)^\ell},
    \label{solution-extremal-H1-Lge2}\\
K^{(1)} &=& A_2\, \frac{\ell[(2\ell+3)x-(\ell+2)]}{2(2\ell+3)x(x-1)^{\ell+2}} + \frac{B_2}{2x(x-1)^\ell}.
    \label{solution-extremal-K1-Lge2}
  \end{eqnarray}
\end{subequations}
The first term of each of Eqs.~\eqref{solution-extremal-H1-Lge2}
and \eqref{solution-extremal-K1-Lge2} behaves as $O(r^{\ell+2})$,
and may be interpreted as the perturbation induced by
addition of the electric multipole moments. By contrast,
the second term of each of Eqs.~\eqref{solution-extremal-H1-Lge2}
and \eqref{solution-extremal-K1-Lge2} behaves as $O(r^{\ell+1})$,
and may be interpreted as the distortion caused by adding
the mass multipole moments since this power appears in the distortion of
a Schwarzschild spacetime as well \cite{Regge:1957}
(see also \cite{Yoshino:2016}). 
However, there is a subtlety in 
dividing the perturbation into the two modes, since there is a freedom
to add a part of the first term to the second term without changing the
primary behavior at the distant place.
This is a difficult problem of interpretation,
and we do not go into further details here.
The electrostatic potential is given by
\begin{equation}
  \varphi^{(1)} \ = \ -A_2\,\frac{(\ell+2)[(2\ell+3)x(x-1)+\ell]}{2(2\ell+3)x^2(x-1)^{\ell+1}}+\frac{B_2}{2x^2(x-1)^{\ell-1}}.
\end{equation}

\subsection{Sub-extremal case}
\label{Sec:SubExtremal-Lge2}

In the sub-extremal case,
following \cite{Bini:2006},
we introduce the coordinate
\begin{equation}
  z:= \frac{r-M}{\gamma M},
  \label{Def:z}
\end{equation}
where $\gamma$ is defined in Eq.~\eqref{Def:Gamma}.
In this coordinate, the two horizons $r=r_{\pm}$ are located
at $z=\pm 1$. We also introduce the function $w(z)$ by
\begin{equation}
  X\,=\,\frac{\gamma M}{re^{\nu^{(0)}}}\ w.
  \label{Introduction-w-subextremal}
\end{equation}
Eq.~\eqref{Equation-for-X} is rewritten as
\begin{equation}
(1-z^2)w_{,zz}-2zw_{,z}+\left[\ell(\ell+1)-\frac{1}{1-z^2}\right]w \, = \, 0.
\end{equation}
This equation is satisfied by the
associated Legendre functions
of the first and second kinds, $P_{\ell}^{\mu}(z)$ and $Q_{\ell}^{\mu}(z)$,
with $\mu=1$. 
Note that the definition of the associated Legendre functions
depends on the authors and the contexts. Here, we adopt the
expressions,
\begin{subequations}
\begin{eqnarray}
  P_{\ell}^{\mu}(z) &=&
  p_{\ell\mu}\left(z^2-1\right)^{\mu/2}
       {}_2F_1\left(-\ell+\mu,\ell+\mu+1;\ell+\mu;\frac{1-z}{2}\right),\quad
       \label{LegendreP-standard}\\
  Q_{\ell}^{\mu}(z) &=& q_{\ell\mu}
  \frac{(z^2-1)^{\mu/2}}{z^{\ell+\mu+1}}
       {}_2F_1\left(\frac{\ell+\mu+2}{2},\frac{\ell+\mu+1}{2};\ell+\frac32;\frac{1}{z^2}\right),\label{LegendreQ-standard}
\end{eqnarray}
\end{subequations}
with
\begin{subequations}
  \begin{eqnarray}
    p_{\ell\mu} &=& \frac{\Gamma(\ell+\mu+1)}{2^\mu\Gamma(\ell-\mu+1)\Gamma(\ell+\mu)},\\
    q_{\ell\mu} &=& \frac{(-1)^\mu\sqrt{\pi}\Gamma(\ell+\mu+1)}{2^{\ell+1}\Gamma(\ell+3/2)},\label{q_ellmu}
\end{eqnarray}
\end{subequations}
where ${}_2F_1(a,b;c;z)$ is the Gauss hypergeometric function.
Here, the formula for $P_{\ell}^{\mu}(z)$ is applicable only to
the case that $\ell$ and $\mu$ are integers.
In Appendix~\ref{App:Explicit-P-Q},
we list up $P_{\ell}^{1}(z)$ and $Q_{\ell}^{1}(z)$
for $\ell=1$, $2$, and $3$.
The two independent solutions for $w(z)$ are 
$P_{\ell}^{1}(z)$ and $Q_{\ell}^{1}(z)$, and
hence, the two independent solutions for $X(z)$ are
$X_1=P_{\ell}^{1}(z)/\sqrt{z^2-1}$ and
\begin{equation}
  X_2 \, = \, \frac{1}{q_{\ell 1}\gamma^{\ell+2}}\frac{Q_{\ell}^{1}(z)}{\sqrt{z^2-1}},
  \label{X2-subextremal-Lge2}
\end{equation}
where the coefficient is chosen so that Eq.~\eqref{X2-subextremal-Lge2}
is reduced to Eq.~\eqref{X2-extremal-Lge2} in the extremal limit $\gamma\to 0$.
Then, $X=A_2X_2$ is adopted as the solution for $X$.

In order to solve for the function $Y$, we introduce
$v$ by
\begin{equation}
  Y \, = \, e^{\nu^{(0)}} v.
  \label{Def-v-subextremal}
\end{equation}
The equation for $v$ becomes
\begin{equation}
  (1-z^2)v_{,zz}-2zv_{,z}
  +\left[\ell(\ell+1)-\frac{1}{1-z^2}\right]v \ = \ A_2\mathcal{S},
  \label{Equation-for-v}
\end{equation}
with
\begin{eqnarray}
  \mathcal{S}& :=& -e^{-\nu^{(0)}}S(X_2) \nonumber
  \\
  & = & \frac{2}{q_{\ell 1}\gamma^{\ell+2}}\frac{z+\gamma^{-1}}{z^2-1}
  \left\{
  \left(z+\frac{z+\gamma}{\gamma z+1}\right)Q_{\ell,z}^1(z)
  -\left[\ell^2+\ell+\frac{1}{\gamma z+1}
    -\frac{2}{1-z^2}\right]Q_{\ell}^1(z)
  \right\}.
\end{eqnarray}
The two independent homogeneous solutions for $v$
are given by $P_{\ell}^1(z)$ and $Q_{\ell}^1(z)$,
and hence, we have
$Y_1 = e^{\nu^{(0)}}P_{\ell}^1(z)$ and
\begin{equation}
  Y_2\, =\, \frac{1}{q_{\ell 1}\gamma^{\ell+1}}\frac{\sqrt{z^2-1}}{z+\gamma^{-1}}\, Q_{\ell}^1(z).
  \label{Y2-subextremal-Lge2}
\end{equation}
Note that Eq.~\eqref{Y2-subextremal-Lge2}
is reduced to Eq.~\eqref{Y2-extremal-Lge2} in the extremal limit $\gamma\to 0$.
Then, $B_2Y_2$ is adopted as the homogeneous solution.

In order to calculate the particular solution,
we apply the Green's function method 
to Eq.~\eqref{Equation-for-v}.
Let us consider  
the Green's function $G_{\ell \mu }(z,z^\prime)$ 
that satisfies 
\begin{equation}
(1-z^2)G_{\ell \mu ,zz}-2zG_{\ell \mu ,z}
  +\left[\ell(\ell+1)-\frac{\mu ^2}{1-z^2}\right]G_{\ell \mu }
  \ = \ \delta(z-z^\prime).
\end{equation}
Since the homogeneous solutions are $P_{\ell}^\mu (z)$
and $Q_{\ell}^\mu (z)$, the method of
constructing $G_{\ell \mu }(z,z^\prime)$ is straightforward:
\begin{equation}
  G_{\ell \mu }(z,z^\prime)
  \ = \
  \frac{1}{W_{\ell \mu }}\left[
P_{\ell}^\mu (z)
Q_{\ell}^\mu (z^\prime)
\theta(z^\prime-z)
+
P_{\ell}^\mu (z^\prime)
Q_{\ell}^\mu (z)
\theta(z-z^\prime)
    \right],
\end{equation}
where $W_{\ell\mu}$ is the Wronskian of $P_{\ell}^\mu (z)$ and $Q_{\ell}^\mu (z)$
multiplied by $(1-z^2)$, i.e.
\begin{eqnarray}
  W_{\ell \mu } &=& (1-z^2)\left[
    P_{\ell}^\mu (z)\frac{Q_{\ell}^\mu (z)}{dz}
    -
    Q_{\ell}^\mu (z)\frac{P_{\ell}^\mu (z)}{dz}
    \right]
  \nonumber\\
  & = &
  \frac{(-1)^\mu  2^{2\mu }\Gamma\left(\frac{\ell+\mu +2}{2}\right)\Gamma\left(\frac{\ell+\mu +1}{2}\right)}{\Gamma\left(\frac{\ell-\mu +2}{2}\right)\Gamma\left(\frac{\ell-\mu +1}{2}\right)},
\end{eqnarray}
which is a constant. The particular solution has
the form $A_2v_{\rm pt}$, where $v_{\rm pt}$ is
calculated by
\begin{eqnarray}
v_{\rm pt}(z) &=& \int_{z_{\rm LB}}^\infty \mathcal{S}(z^\prime)G_{\ell 1}(z,z^\prime)dz^\prime
\nonumber\\
&=&
\frac{1}{W_{\ell 1}}\left[P_{\ell}^1(z)\int_{z}^\infty\mathcal{S}(z^\prime)Q_{\ell}^1(z^\prime)dz^\prime
  +
Q_{\ell}^1(z)\int_{z_{\rm LB}}^z\mathcal{S}(z^\prime)P_{\ell}^1(z^\prime)dz^\prime 
  \right].
\end{eqnarray}
Here, the lower bound of the integration, $z_{\rm LB}$,
can be arbitrarily chosen, because changing $z_{\rm LB}$
corresponds to adding the homogeneous solution
to the particular solution.
We choose $z_{\rm LB}=\infty$ so that $Y_{\rm pt}$ behaves
as $Y_{\rm pt}\sim 1/r^{\ell+2}$ at large $r$. Then,
the particular solution $Y_{\rm pt}$ is expressed as
\begin{equation}
  Y_{\rm pt} \ = \ \frac{1}{W_{\ell 1}}\,\frac{\sqrt{z^2-1}}{z+\gamma^{-1}}
  \left[Q_{\ell}^1(z)I^{(P)}_{\ell 1}(z) - P_{\ell}^1(z)I^{(Q)}_{\ell 1}(z)\right],
  \label{Ypt-subextremal-Lge2}
\end{equation}
where
\begin{subequations}
\begin{eqnarray}
I^{(P)}_{\ell 1}(z) &=& \int_{\infty}^z\mathcal{S}(z^\prime)P_{\ell}^1(z^\prime)dz^\prime,\\
I^{(Q)}_{\ell 1}(z) &=& \int_{\infty}^z\mathcal{S}(z^\prime)Q_{\ell}^1(z^\prime)dz^\prime,
\end{eqnarray}
\end{subequations}
and the formal solution of $Y$ in the form of Eq.~\eqref{formal-solution-Y}
is given. 
Although we could not find general formulas for
$I^{(P)}_{\ell 1}(z)$ and $I^{(Q)}_{\ell 1}(z)$, it is possible to
calculate it using, e.g. {\it Mathematica} for each value of $\ell$.
The formulas for $Y_{\rm pt}$ obtained in this way are presented
in Appendix~\ref{App:SubExtremal-Ypt} for $\ell=2$ and $3$.
Using the formulas presented in Appendix~\ref{App:Explicit-P-Q}
and \ref{App:SubExtremal-Ypt}, $H^{(1)}$ and $K^{(1)}$ are easily computed.
Since the calculation of $\varphi^{(1)}$ is somewhat tedious,
we present their formulas for $\ell=2$ and $3$ in Appendix~\ref{App:SubExtremal-varphi}.

\subsection{Super-extremal case}
\label{Sec:SuperExtremal-Lge2}

In the super-extremal case,
introducing the $z$ coordinate with Eq.~\eqref{Def:z}
and the function $w(z)$ with Eq.~\eqref{Introduction-w-subextremal},
Eq.~\eqref{Equation-for-X} is rewritten as
\begin{equation}
(1+z^2)w_{,zz}+2zw_{,z}-\left[\ell(\ell+1)-\frac{1}{1+z^2}\right]w \, = \, 0.
\end{equation}
Introducing the coordinate $\zeta=iz$, the equation for
$w$ becomes the equation for the associated Legendre functions,
\begin{equation}
(1-\zeta^2)w_{,\zeta\zeta}-2\zeta w_{,\zeta}+\left[\ell(\ell+1)-\frac{1}{1-\zeta^2}\right]w \, = \, 0.
\end{equation}
This means that the two independent solutions for $w(z)$
is ${P}_{\ell}^{1}(iz)$ and $Q_{\ell}^1(iz)$.
For simplicity, we introduce the real-valued functions,
$\tilde{P}_{\ell}^{\mu}(z)=(i)^{-\ell}{P}_{\ell}^{\mu}(iz)$ and
$\tilde{Q}_{\ell}^{\mu}(z)=(i)^{2\mu+\ell+1}{Q}_{\ell}^{\mu}(iz)$, or
equivalently, 
\begin{subequations}
\begin{eqnarray}
  \tilde{P}_{\ell}^{\mu}(z) &=&
  \tilde{p}_{\ell\mu}\left(z^2+1\right)^{\mu/2}
       {}_2F_1\left(-\ell+\mu,\ell+\mu+1;\ell+\mu;\frac{1-iz}{2}\right),\quad
       \label{Def:TildeP}\\
  \tilde{Q}_{\ell}^{\mu}(z) &=&
  \tilde{q}_{\ell\mu}\frac{(z^2+1)^{\mu/2}}{z^{\ell+\mu+1}}
       {}_2F_1\left(\frac{\ell+\mu+2}{2},\frac{\ell+\mu+1}{2};\ell+\frac32;-\frac{1}{z^2}\right),\label{Def:TildeQ}
\end{eqnarray}
\end{subequations}
where
\begin{subequations}
  \begin{eqnarray}
\tilde{p}_{\ell\mu} &=& i^{\mu-\ell}p_{\ell\mu},\\
\tilde{q}_{\ell\mu} &=& (-1)^{\mu}q_{\ell\mu}.\label{tilq_ellmu}
  \end{eqnarray}
\end{subequations}
The explicit formulas for $\tilde{P}_{\ell}^{1}(z)$
and $\tilde{Q}_{\ell}^{1}(z)$ are presented in Appendix~\ref{App:Explicit-tilP-tilQ} for $\ell=1$, $2$, and $3$.
The two independent solutions for $X(z)$ are
$X_1=\tilde{P}_{\ell}^{1}(z)/\sqrt{z^2+1}$ and
\begin{equation}
  X_2 \, = \, \frac{1}{\tilde{q}_{\ell 1}\gamma^{\ell+2}}\frac{\tilde{Q}_{\ell}^{1}(z)}{\sqrt{z^2+1}}.
  \label{X2-superextremal-Lge2}
\end{equation}
Equation~\eqref{X2-superextremal-Lge2} reduces to Eq.~\eqref{X2-extremal-Lge2}
in the extremal limit $\gamma\to 0$. Then,
$X=A_2X_2$ is adopted as the solution for $X$.

In order to solve for $Y$, we introduce the function $v$
with Eq.~\eqref{Def-v-subextremal}.
The equation for $v$ becomes
\begin{equation}
  (1+z^2)v_{,zz}+2zv_{,z}
  -\left[\ell(\ell+1)-\frac{1}{1+z^2}\right]v \ = \ A_2\tilde{\mathcal{S}},
  \label{Equation-for-v-superextremal}
\end{equation}
with
\begin{eqnarray}
  \tilde{\mathcal{S}}& :=& e^{-\nu^{(0)}}S(X_2) \nonumber
  \\
  & = & -\frac{2}{\tilde{q}_{\ell 1}\gamma^{\ell+2}}\frac{z+\gamma^{-1}}{z^2+1}
  \left\{
  \left(z+\frac{z-\gamma}{\gamma z+1}\right)\tilde{Q}_{\ell,z}^1(z)
  -\left[\ell^2+\ell+\frac{1}{\gamma z+1}
    -\frac{2}{z^2+1}\right]\tilde{Q}_{\ell}^1(z)
  \right\}.
\end{eqnarray}
The two independent homogeneous solutions for $v$
are given by $\tilde{P}_{\ell}^1(z)$ and $\tilde{Q}_{\ell}^1(z)$,
and hence, we have
$Y_1 = e^{\nu^{(0)}}\tilde{P}_{\ell}^1(z)$ and
\begin{equation}
  Y_2\, =\, \frac{1}{\tilde{q}_{\ell 1}\gamma^{\ell+1}}\frac{\sqrt{z^2+1}}{z+\gamma^{-1}}\, \tilde{Q}_{\ell}^1(z),
  \label{Y2-superextremal-Lge2}
\end{equation}
where the coefficient is adjusted to make Eq.~\eqref{Y2-superextremal-Lge2}
consistent with Eq.~\eqref{Y2-extremal-Lge2} in the extremal limit
$\gamma\to 0$.
Then, $B_2Y_2$ is adopted as the homogeneous solution.

In order to calculate the particular solution,
we consider  
the Green's function $G_{\ell \mu }(z,z^\prime)$ 
that satisfies 
\begin{equation}
(1+z^2)\tilde{G}_{\ell \mu ,zz}+2z\tilde{G}_{\ell \mu ,z}
  -\left[\ell(\ell+1)-\frac{\mu ^2}{1+z^2}\right]\tilde{G}_{\ell \mu }
  \ = \ \delta(z-z^\prime).
\end{equation}
Using the homogeneous solutions $\tilde{P}_{\ell}^\mu (z)$
and $\tilde{Q}_{\ell}^\mu (z)$, the Green's function
$G_{\ell \mu }(z,z^\prime)$ is straightforwardly constructed:
\begin{equation}
  \tilde{G}_{\ell \mu }(z,z^\prime)
  \ = \
  \frac{1}{\tilde{W}_{\ell \mu }}\left[
\tilde{P}_{\ell}^\mu (z)
\tilde{Q}_{\ell}^\mu (z^\prime)
\theta(z^\prime-z)
+
\tilde{P}_{\ell}^\mu (z^\prime)
\tilde{Q}_{\ell}^\mu (z)
\theta(z-z^\prime)
    \right],
\end{equation}
where $\tilde{W}_{\ell\mu}$ is the Wronskian of
$\tilde{P}_{\ell}^\mu (z)$ and $\tilde{Q}_{\ell}^\mu (z)$
multiplied by $(1+z^2)$, i.e.
\begin{eqnarray}
  \tilde{W}_{\ell \mu } &=& (1+z^2)\left[
    \tilde{P}_{\ell}^\mu (z)\frac{\tilde{Q}_{\ell}^\mu (z)}{dz}
    -
    \tilde{Q}_{\ell}^\mu (z)\frac{\tilde{P}_{\ell}^\mu (z)}{dz}
    \right]
  \ = \ (-1)^{\mu+1}W_{\ell\mu},
\end{eqnarray}
which is a constant.
Similarly to the sub-extremal case, the particular solution
is formally expressed as
\begin{equation}
  Y_{\rm pt} \ = \ \frac{1}{\tilde{W}_{\ell 1}}\,\frac{\sqrt{z^2+1}}{z+\gamma^{-1}}
  \left[\tilde{Q}_{\ell}^1(z)\tilde{I}^{(\tilde{P})}_{\ell 1}(z)
    - \tilde{P}_{\ell}^1(z)I^{(\tilde{Q})}_{\ell 1}(z)\right],
  \label{Ypt-superextremal-Lge2}
\end{equation}
where
\begin{subequations}
\begin{eqnarray}
\tilde{I}^{(\tilde{P})}_{\ell 1}(z) &=& \int_{\infty}^z\tilde{\mathcal{S}}(z^\prime)\tilde{P}_{\ell}^1(z^\prime)dz^\prime,\\
\tilde{I}^{(\tilde{Q})}_{\ell 1}(z) &=& \int_{\infty}^z\tilde{\mathcal{S}}(z^\prime)\tilde{Q}_{\ell}^1(z^\prime)dz^\prime.
\end{eqnarray}
\end{subequations}
Then, the formal solution of $Y$ in the form of Eq.~\eqref{formal-solution-Y}
is given. Although we could not find general formulas for
$\tilde{I}^{(\tilde{P})}_{\ell 1}(z)$ and $\tilde{I}^{(\tilde{Q})}_{\ell 1}(z)$,
it is possible to
calculate them using {\it Mathematica} by specifying the value of $\ell$.
The formulas for $Y_{\rm pt}$ obtained in this way are presented
in Appendix~\ref{App:SuperExtremal-Ypt} for $\ell=2$ and $3$.
Since the calculation of $\varphi^{(1)}$ is somewhat tedious,
we present their formulas for $\ell=2$ and $3$ in Appendix~\ref{App:SuperExtremal-varphi}.

%
%
\section{Solving the perturbative equations for $\ell= 1$}
\label{Sec:Solving_equations_L1}

We now study the case $\ell=1$.
As discussed in Sec.~\ref{Sec:Equation-for-L1},
it is possible to require the gauge condition
$H^{(1)}=L^{(1)}$, and we have to solve
the same equations for $X$ and $Y$,
Eqs.~\eqref{Equation-for-X} and \eqref{Equation-for-Y} by setting $\ell=1$.
Here, we have to be careful because there are
several differences from the $\ell\ge 2$ case.
First, the both of the two independent solution
$X_1$ and $X_2$ of Eq.~\eqref{Equation-for-X}
do not diverge at $r\to\infty$
because they behave as $X_1\to r^{0}$ and $X\sim 1/r^3$.
Therefore, both of them can be
adopted as the solutions.
Here, $X_1$ is found to be a constant, and the solution can be written
as
\begin{equation}
  X\, = \, A_1\,+\,A_2\,X_2.
  \label{formal-solution-X-L1}
\end{equation}
Then, the equation for $Y$ is written as
\begin{equation}
  r^2e^{2\nu^{(0)}} \left(Y_{,rr}+\frac{2}{r}Y_{,r}\right)
  -2Y
  \, = \,
  2A_1+A_2S(X_2).
  \label{Equation-for-Y-L1}
\end{equation}
The solution for $Y$ is the sum of the homogeneous solution
and the particular solution. 
The two independent homogeneous solution $Y_1$ and $Y_2$
behave as $Y_1\sim r$ and $Y_2\sim 1/r^2$ at large $r$,
and only the solution $Y_2$ is appropriate in our setup.
Since there are two sources $2A_1$ and $A_2S(X_2)$
in Eq.~\eqref{Equation-for-Y-L1}, the particular solution
can be written as $A_1Y_{\rm pt}^{(1)}+A_2Y_{\rm pt}^{(2)}$.
Here, it is easily checked that $Y_{\rm pt}^{(1)}=-1$ holds 
by substituting into Eq.~\eqref{Equation-for-Y-L1}.
In total, the solution for Eq.~\eqref{Equation-for-Y-L1}
is written as
\begin{equation}
  Y \ = \ B_2Y_2 \, - \, A_1 \,+\, A_2Y_{\rm pt},
  \label{formal-solution-Y-L1}
\end{equation}
where we have set $Y_{\rm pt}=Y_{\rm pt}^{(2)}$. 
Then, we can write down $H^{(1)}$ and $K^{(1)}$ using Eqs.~\eqref{Def:X}
and \eqref{Def:Y}.
Although the equation for $H^{(1)}$ is the same as
Eq.~\eqref{calculate-H1-from-X-Y}, the equation for $K^{(1)}$
is modified as
\begin{equation}
K^{(1)} \ =\ -A_1+A_2\frac{Y_{\rm pt}-X_2}{2} + B_2\frac{Y_2}{2}.
  \label{calculate-K1-from-X-Y-L1}
\end{equation}
$\varphi^{(1)}$ is calculated
with Eq.~\eqref{Rr theta}.

The above solution includes the gauge-mode solution,
Eqs.~\eqref{gauge-mode-solution-H1}, \eqref{gauge-mode-solution-K1}
and \eqref{gauge-mode-solution-phi1} 
with $b(r)$ given by Eq.~\eqref{gauge-mode-function-b},
as discussed in Sec.~\ref{Sec:Equation-for-L1}.
For this reason, it must be possible to rewrite the obtained solution
into the form
\begin{subequations}
\begin{eqnarray}
  H^{(1)} & = & -\frac12r^2\left(e^{2\nu^{(0)}}\right)_{,r}b_{,r}
  +D\, h^{(1)}_{\rm phys},\label{formal-solution-H1-L1}\\
  K^{(1)} & = & re^{2\nu^{(0)}}b_{,r}+b
  +D\, k^{(1)}_{\rm phys}.\label{formal-solution-K1-L1}
\end{eqnarray}
\end{subequations}
Then, $H^{(1)}=D\, h^{(1)}_{\rm phys}$ and $K^{(1)}=D\, k^{(1)}_{\rm phys}$
are adopted as the physical perturbation
after checking that they are not proportional to the gauge-mode solutions.
Note that in the Schwarzschild case,
Eq.~\eqref{Rr theta} plays a role to give a constraint for the
functions $H^{(1)}$ and $K^{(1)}$ rather than determining $\varphi^{(1)}$,
and thus, only the gauge mode is present.
This causes the difference between the Reissner-Nordstr\"om case
and the Schwarzschild case. Physically, 
in the Schwarzschild case, only the gauge mode appears in the $\ell=1$ mode 
because addition of mass dipole moment is equivalent to shifting the 
central mass position, and is absorbed by the gauge transformation. 
The physical mode appears in the 
Reissner-Nordstr\"om case because adding an electric dipole moment
changes the energy-momentum tensor, and hence,
causes the actual change in the gravitational field.

Below, we present the functions $X_2$, $Y_2$, 
$Y_{\rm pt}$, $h^{(1)}_{\rm phys}$, and $k^{(1)}_{\rm phys}$
for the extremal, sub-extremal, and super-extremal cases, one by one.

\subsection{Extremal case}
\label{Sec:Extremal-Leq1}

Using the rescaled radial coordinate $x$ defined in Eq.~\eqref{Def:x},
the same formulas for $X_2$, $Y_2$, and $Y_{\rm pt}$
as the cases $\ell\ge 2$, Eqs.~\eqref{X2-extremal-Lge2}, \eqref{Y2-extremal-Lge2}, and \eqref{Ypt-extremal-Lge2}, 
can be used by substituting $\ell=1$:
\begin{eqnarray}
  X_2 &=& \frac{1}{(x-1)^3},\\
  Y_2 &=& \frac{1}{x(x-1)},\\
  Y_{\rm pt} &=& \frac{2}{x(x-1)^2}+\frac{7}{5x(x-1)^3}.
\end{eqnarray}
Subsutituting into Eqs.~\eqref{formal-solution-X-L1}
and \eqref{formal-solution-Y-L1}, we have the solution for $X$ and
$Y$, and hence, the solution for $H^{(1)}$ and $K^{(1)}$.
Setting $A_1=-\alpha$, $A_2=D$, and $B_2=2\beta$,
we have
\begin{subequations}
\begin{eqnarray}
  H^{(1)} &=& \frac{\beta}{x(x-1)} + D\,\frac{3(5x-1)}{10x(x-1)^3},
  \label{Explicit-solution-extremal-H1-L1}\\
  K^{(1)} &=& \alpha+\frac{\beta}{x(x-1)} + D\,\frac{5x-3}{10x(x-1)^3}.
  \label{Explicit-solution-extremal-K1-L1}
\end{eqnarray}
\end{subequations}
The first term of the right-hand side of Eq.~\eqref{Explicit-solution-extremal-H1-L1} agrees with that of Eq.~\eqref{formal-solution-H1-L1},
and the first two terms of the right-hand side of Eq.~\eqref{Explicit-solution-extremal-K1-L1} coincide with those of Eq.~\eqref{formal-solution-K1-L1},
using the formula for $b(r)$ given in Eq.~\eqref{gauge-mode-function-b}.
Therefore, we can determine $h^{(1)}_{\rm phys}$ and $k^{(1)}_{\rm phys}$ 
as
\begin{subequations}
\begin{eqnarray}
  h^{(1)}_{\rm phys} &=& \frac{3(5x-1)}{10x(x-1)^3},
  \label{Physical-part-extremal-H1-L1}\\
  k^{(1)}_{\rm phys} &=& \frac{5x-3}{10x(x-1)^3}.
  \label{Physical-part-extremal-K1-L1}
\end{eqnarray}
\end{subequations}
The electrostatic potential is
\begin{equation}
  \varphi^{(1)} \ = \ \frac{\beta}{x^2}-D\,\frac{3[5x(x-1)+1]}{10x^2(x-1)^2}.
  \label{varphi_L1_extremal}
\end{equation}
The first term corresponds to Eq.~\eqref{gauge-mode-solution-phi1}
and the second term represents the physical part.

\subsection{Sub-extremal case}
\label{Sec:SubExtremal-Leq1}

Using the coordinate $z$ defined in Eq.~\eqref{Def:z},
the same formulas for $X_2$, $Y_2$, and $Y_{\rm pt}$
as the cases $\ell\ge 2$, Eqs.~\eqref{X2-subextremal-Lge2}, \eqref{Y2-subextremal-Lge2}, and \eqref{Ypt-subextremal-Lge2}, 
can be used by substituting $\ell=1$.
Using the formulas for $P_{1}^1(z)$ and $Q_1^1(z)$
presented in Appendix~\ref{App:Explicit-P-Q}, we find
\begin{eqnarray}
  X_2 &=& \frac{3}{2\gamma^3}\left[\frac{z}{z^2-1}-\mathrm{arccoth}(z)\right],\\
  Y_2 &=& \frac{3}{2\gamma(\gamma z+1)}\left[z-(z^2-1)\mathrm{arccoth}(z)\right],\\
  Y_{\rm pt} &=& -\frac{3}{2\gamma^3(\gamma z+1)}\left[\frac{3z^3+\gamma z^2-4z-2\gamma}{z^2-1}-{(3z^2+\gamma z-2)}\mathrm{arccoth}(z)\right].
\end{eqnarray}
Subsutituting into Eqs.~\eqref{formal-solution-X-L1}
and \eqref{formal-solution-Y-L1}, we have the solution for $X$ and
$Y$, and hence, the solution for $H^{(1)}$ and $K^{(1)}$.
Setting $A_1=-\alpha$, $A_2=(2/3)\gamma^2\beta+D$, and $B_2=2\beta$,
we have
\begin{subequations}
\begin{eqnarray}
  H^{(1)} &=& \frac{\beta}{\gamma}\,\frac{z+\gamma}{(z^2-1)(\gamma z+1)}
  + \frac{3D}{4\gamma^3(\gamma z+1)}\,
  \left[-\frac{3z^3-5z-2\gamma}{z^2-1}+3(z^2-1)\mathrm{arccoth}(z)\right], \qquad
  \label{Explicit-solution-subextremal-H1-L1}\\
  K^{(1)} &=& \alpha+\beta\,\left[\frac{-1}{\gamma z+1}+
    \frac{\mathrm{arccoth}(z)}{\gamma}\right]
  \nonumber\\
  && \qquad\qquad\qquad\qquad\quad~
  - \frac{3D}{4\gamma^3(\gamma z+1)}\,
  \left[3z+2\gamma-(3z^2+2\gamma z-1)\mathrm{arccoth}(z)\right].
  \label{Explicit-solution-subextremal-K1-L1}
\end{eqnarray}
\end{subequations}
The first term of the right-hand side of Eq.~\eqref{Explicit-solution-subextremal-H1-L1} agrees with that of Eq.~\eqref{formal-solution-H1-L1},
and the first two terms of the right-hand side of Eq.~\eqref{Explicit-solution-subextremal-K1-L1} coincide with those of Eq.~\eqref{formal-solution-K1-L1},
using the formula for $b(r)$ given in Eq.~\eqref{gauge-mode-function-b}.
Therefore, we can determine $h^{(1)}_{\rm phys}$ and $k^{(1)}_{\rm phys}$ 
as
\begin{subequations}
\begin{eqnarray}
  h^{(1)}_{\rm phys} & = & \frac{3}{4\gamma^3(\gamma z+1)}\,
  \left[-\frac{3z^3-5z-2\gamma}{z^2-1}+3(z^2-1)\mathrm{arccoth}(z)\right],
  \label{Physical-part-subextremal-H1-L1}
  \\
  k^{(1)}_{\rm phys} & = & - \frac{3}{4\gamma^3(\gamma z+1)}\,
  \left[3z+2\gamma-(3z^2+2\gamma z-1)\mathrm{arccoth}(z)\right].
  \label{Physical-part-subextremal-K1-L1}
\end{eqnarray}
\end{subequations}
Note that at large $z$, these functions
behave as $h^{(1)}_{\rm phys}\approx 3/2x^3$ and
$k^{(1)}_{\rm phys}\approx 1/2x^3$ where $x$ is defined in Eq.~\eqref{Def:x}.
Also, 
the functions $H^{(1)}$ and $K^{(1)}$ of Eqs.~\eqref{Explicit-solution-subextremal-H1-L1}
and \eqref{Explicit-solution-subextremal-K1-L1}
reduce to those of the extremal case, Eqs.~\eqref{Explicit-solution-extremal-H1-L1} and \eqref{Explicit-solution-extremal-K1-L1},
in the limit $\gamma\to 0$.
The electrostatic potential is
\begin{equation}
  \varphi^{(1)} \ =\
  \frac{\beta\sqrt{1-\gamma^2}}{(\gamma z+1)^2}
  -D\frac{3[3z(z+\gamma)-2(1-\gamma^2)-3(z^2-1)(z+\gamma)\mathrm{arccoth}(z)]}{4\gamma^2\sqrt{1-\gamma^2}(\gamma z+1)^2}.
  \label{varphi_L1_subextremal}
\end{equation}
The first term corresponds to Eq.~\eqref{gauge-mode-solution-phi1}
and the second term represents the physical part.

\subsection{Super-extremal case}
\label{Sec:SuperExtremal-Leq1}

Finally, we consider the super-extremal case.
Using the coordinate $z$ defined in Eq.~\eqref{Def:z},
the same formulas for $X_2$, $Y_2$, and $Y_{\rm pt}$
as the cases $\ell\ge 2$, Eqs.~\eqref{X2-superextremal-Lge2}, \eqref{Y2-superextremal-Lge2}, and \eqref{Ypt-superextremal-Lge2}, 
can be used by substituting $\ell=1$.
Using the formulas for $\tilde{P}_{1}^1(z)$ and $\tilde{Q}_1^1(z)$
presented in Appendix~\ref{App:Explicit-P-Q}, we find
\begin{eqnarray}
  X_2 &=& \frac{3}{2\gamma^3}\left[-\frac{z}{z^2+1}+\mathrm{arccot}\left(z\right)\right],\\
  Y_2 &=& -\frac{3}{2\gamma(\gamma z+1)}\left[z-(z^2+1)\mathrm{arccot}\left(z\right)\right],\\
  Y_{\rm pt} &=& \frac{3}{2\gamma^3}\left[\frac{-3z^3+\gamma z^2-4z+2\gamma}{(z^2+1)(\gamma z+1)}+\frac{3z^2-\gamma z+2}{\gamma z+1}\mathrm{arccot}\left(z\right)\right].
\end{eqnarray}
Subsutituting into Eqs.~\eqref{formal-solution-X-L1}
and \eqref{formal-solution-Y-L1}, we have the solution for $X$ and
$Y$, and hence, the solution for $H^{(1)}$ and $K^{(1)}$.
Setting $A_1=-\alpha$, $A_2=-(2/3)\gamma^2\beta+D$, and $B_2=2\beta$,
we have
\begin{subequations}
\begin{eqnarray}
  H^{(1)} &=& \frac{\beta}{\gamma}\,\frac{z-\gamma}{(z^2+1)(\gamma z+1)}
  + \frac{3D}{4\gamma^3(\gamma z+1)}\,
  \left[\frac{-3z^3-5z+2\gamma}{z^2+1}+3(z^2+1)\mathrm{arccot}(z)\right], \qquad
  \label{Explicit-solution-superextremal-H1-L1}\\
  K^{(1)} &=& \alpha+\frac{\beta}{\gamma}\,\left[-\frac{\gamma}{\gamma z+1}+\mathrm{arccot}(z)\right]
  \nonumber\\
  && \qquad\qquad\qquad\quad
  + \frac{3D}{4\gamma^3(\gamma z+1)}\,
  \left[-3z+2\gamma+(3z^2-2\gamma z+1)\mathrm{arccot}(z)\right].
  \label{Explicit-solution-superextremal-K1-L1}
\end{eqnarray}
\end{subequations}
The first term of the right-hand side of Eq.~\eqref{Explicit-solution-subextremal-H1-L1} agrees with that of Eq.~\eqref{formal-solution-H1-L1},
and the first two terms of the right-hand side of Eq.~\eqref{Explicit-solution-subextremal-K1-L1} coincide with those of Eq.~\eqref{formal-solution-K1-L1},
using the formula for $b(r)$ given in Eq.~\eqref{gauge-mode-function-b}.
Therefore, we can determine $h^{(1)}_{\rm phys}$ and $k^{(1)}_{\rm phys}$ 
as
\begin{subequations}
\begin{eqnarray}
  h^{(1)}_{\rm phys} & = & \frac{3}{4\gamma^3(\gamma z+1)}\,
  \left[\frac{-3z^3-5z+2\gamma}{z^2+1}+3(z^2+1)\mathrm{arccot}(z)\right],
  \label{Physical-part-superextremal-H1-L1}
  \\
  k^{(1)}_{\rm phys} & = &  \frac{3}{4\gamma^3(\gamma z+1)}\,
  \left[-3z+2\gamma+(3z^2-2\gamma z+1)\mathrm{arccot}(z)\right].
  \label{Physical-part-superextremal-K1-L1}
\end{eqnarray}
\end{subequations}
Note that at large $z$, these functions
behave as $h^{(1)}_{\rm phys}\approx 3/2x^3$ and
$k^{(1)}_{\rm phys}\approx 1/2x^3$.
Also, the functions $H^{(1)}$ and $K^{(1)}$ of Eqs.~\eqref{Explicit-solution-superextremal-H1-L1}
and \eqref{Explicit-solution-superextremal-K1-L1}
reduce to those of the extremal case, Eqs.~\eqref{Explicit-solution-extremal-H1-L1} and \eqref{Explicit-solution-extremal-K1-L1},
in the limit $\gamma\to 0$ in the range $x>1$.
The electrostatic potential is
\begin{equation}
  \varphi^{(1)} \ =\
  \frac{\beta \sqrt{1+\gamma^2}}{(\gamma z+1)^2}
  +D\frac{3[-3z(z-\gamma)-2(1+\gamma^2)+3(z^2+1)(z-\gamma)\mathrm{arccot}(z)]}{4\gamma^2\sqrt{1+\gamma^2}(\gamma z+1)^2}.
  \label{varphi_L1_superextremal}
\end{equation}
The first term corresponds to Eq.~\eqref{gauge-mode-solution-phi1}
and the second term represents the physical part.

%
%
\section{Photon surfaces in distorted configurations}
\label{Sec:Distortion-PS}

In this section, we examine whether 
a static photon surface
is possible to exist in a perturbed Reissner-Nordstr\"om spacetime.

\subsection{Photon surface condition}

We begin with presenting the photon surface condition $\chi_{ab}\propto h_{ab}$
in the current setup. 
Suppose 
the position of a static photon surface to be given as 
\begin{equation}
  r=f(\theta,\phi),
  \label{photon-surface-1}
\end{equation}
in the perturbed Reissner-Nordstr\"om spacetime, where
\begin{equation}
  f = f^{(0)} + \epsilon f^{(1)} +\cdots, 
  \label{photon-surface-2}
\end{equation}
with $f^{(0)}=r_{\rm p}$.
Here, $f^{(1)}$ is supposed to be a function of the angular coordinates,
$f^{(1)}(\theta, \phi)$.
The calculation of the photon surface condition,
$\chi_{ab} \propto h_{ab}$, in a perturbed spacetime
with spherically symmetric background 
has been done in our Appendix A
of our previous paper \cite{Yoshino:2016}. The calculation is applicable
to the current setup except the parts 
where the Schwarzschild property is used. The result is
\begin{subequations}
\begin{eqnarray}
  f^{(1)}_{,\theta\phi}&=&\cot\theta f^{(1)}_{,\phi},
  \label{ps-eq1}
  \\
  f^{(1)}_{,\phi\phi}&=&\sin^2\theta f^{(1)}_{,\theta\theta} -\sin\theta\cos\theta f^{(1)}_{,\theta},
  \label{ps-eq2}
\end{eqnarray}
\begin{equation}
  \left.\left(\nu^{(1)}_{,r}-\psi^{(1)}_{,r}\right)\right|_{r=r_{\rm p}}
  =-\left[\left(\nu^{(0)}_{,rr}+\frac{1}{r^2}\right)f^{(1)}-\frac{1}{r^2e^{2\nu^{(0)}}}f^{(1)}_{,\theta\theta}\right]_{r=r_{\rm p}}.
  \label{ps-eq3}
\end{equation}
\end{subequations}
Here, Eq.~\eqref{ps-eq3} has been modified
from Eq.~(22c) of Ref.~\cite{Yoshino:2016} 
because the Schwarzschild setup was used there.

The solution to the first two equations \eqref{ps-eq1} and \eqref{ps-eq2} are 
found as
\begin{equation}
  f^{(1)} = C_{1}^1\, Y_{1}^{1}(\theta,\phi) + C_1^0\,Y_{1}^{0}(\theta)
  +C_1^{-1}\,Y_{1}^{-1}(\theta,\phi) + C_0^0\, Y_{0}^{0},
  \label{shift-expand}
\end{equation}
where $C_{1}^1$, $C_{1}^0$, $C_{1}^{-1}$, and $C_{0}^0$ are integral
constants.
Since this is a linear combination of the
four spherical harmonics $Y_{\ell}^{m}(\theta,\phi)$
with $(\ell,m) = (1,\pm 1)$, $(1,0)$, and $(0,0)$,
the right-hand side of Eq.~\eqref{ps-eq3} has
the $\ell = 0$ and $1$ modes.
Therefore, again, we have to consider the modes $\ell\ge 2$
and the mode $\ell=1$ separately.

\subsection{The case $\ell \ge 2$}

%
\begin{figure}
  \centering
  \includegraphics[width=0.5\textwidth,bb=0 0 294 279]{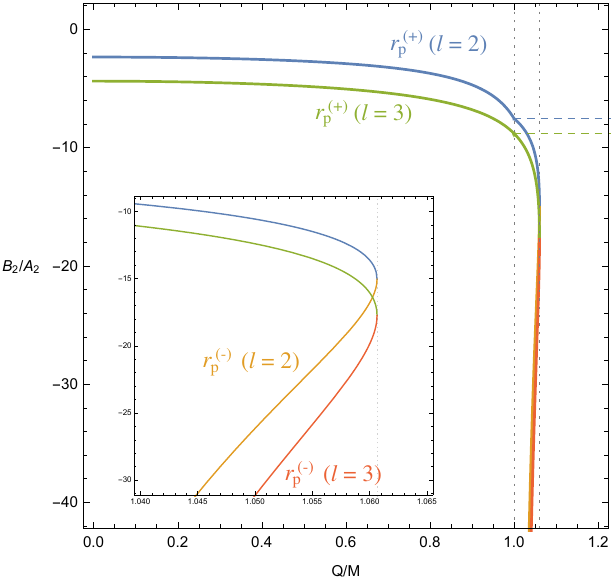}
  \caption{The values of $B_2/A_2$ that realize the distorted
    photon surface at $r_{\rm p}^{(+)}$ and $r_{\rm p}^{(-)}$ for $\ell=2$ and $3$
    as functions of $Q/M$. Each $\ell$ mode has two branches, and
    the upper and lower branches correspond to the outer and
    inner photon surfaces, respectively.
    The left and right vertical dotted lines indicate
    $Q/M=1$ and $\sqrt{9/8}$, respectively. The upper and lower
    horizontal dashed lines indicates the values of $B_2/A_2$
    in the extremal case given by Eq.~\eqref{ratio-B2A2-extremal} 
    for $\ell=2$ and $3$, respectively. The inset
  enlarges the region around $Q/M=\sqrt{9/8}$.}
  \label{Fig:DrawRatioLge2_combined}
\end{figure}
%

%
\begin{figure}
  \centering
  \includegraphics[width=0.45\textwidth,bb=0 0 294 279]{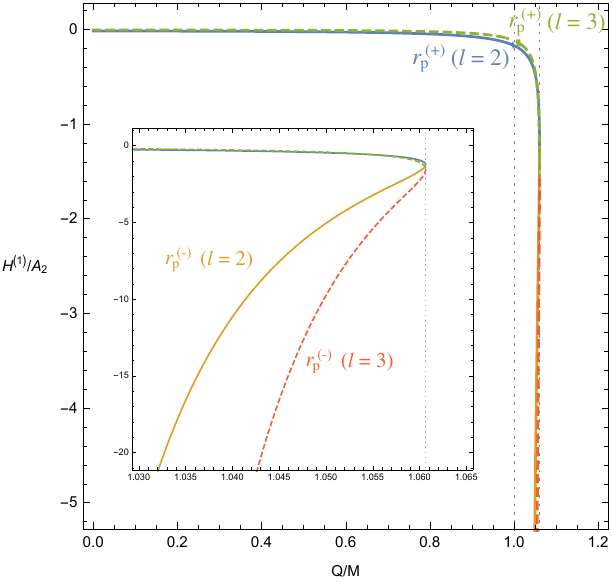}
  \includegraphics[width=0.45\textwidth,bb=0 0 294 276]{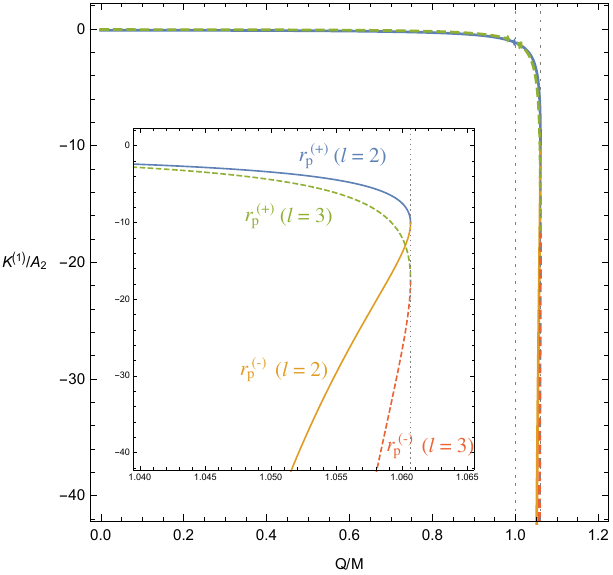}
  \includegraphics[width=0.45\textwidth,bb=0 0 294 278]{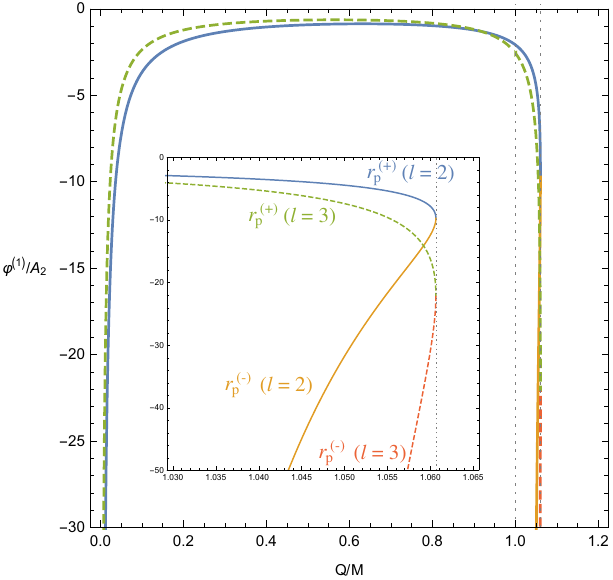}
  \caption{The values of $H^{(1)}/A_2$ (top left), $K^{(1)}/A_2$ (top right), and $\varphi^{(1)}/A_2$ (bottom)
    on the 
    photon surfaces for $\ell=2$ (the solid curve) and $3$ (the dashed curve)
    as functions of $Q/M$. In each panel, each $\ell$ mode has two branches, and
    the upper and lower branches correspond to the outer and
    inner photon surfaces, respectively. The inset
    enlarges the region around $Q/M=\sqrt{9/8}$.
    Since these quantities are nonzero, our results are
    consistent with the uniqueness theorem for the photon sphere,
    and the spatial section of the photon surface is distorted.}
  \label{Fig:Ricci2D_Lge2_combined}
\end{figure}
%

In the cases of $\ell\ge 2$,
the left-hand side of Eq.~\eqref{ps-eq3}
has the $\ell\ge 2$ modes while
the right-hand side only possesses $\ell=0$ and $1$ modes.
Therefore, both sides of Eq.~\eqref{ps-eq3} become zero,
and thus, we have
\begin{equation}
f^{(1)} =0.
\end{equation}
Namely, for the $\ell\ge 2$ modes in the Regge-Wheeler gauge,
the coordinate
position of the distorted photon surface 
must remain at $r=r_{\rm p}$ to first order.
The left-hand side leads to the 
condition for the perturbation quantities,
\begin{equation}
  \left.\left(\nu^{(1)}_{,r}-\psi^{(1)}_{,r}\right)\right|_{r=r_{\rm p}} = 0,
  \label{ps-condition-metric}
\end{equation}
for $r=r_{\rm p}$ to be a photon surface.
From Eqs.~\eqref{function_nu1} and \eqref{function_psi1},
this is equivalent to
\begin{equation}
Y_{,r}(r_{\rm p}) \, = \, 0,
\end{equation}
where $Y$ is defined in Eq.~\eqref{Def:Y}.
From Eq.~\eqref{formal-solution-Y}, this is rewritten as
\begin{equation}
\frac{B_2}{A_2}\, = \, -\left.\frac{Y_{\mathrm{pt},r}}{Y_{2,r}}\right|_{r=r_{\rm p}}.
\label{PS-condition-Lge2}
\end{equation}
In the extremal case,
this condition is written as
\begin{equation}
  \frac{B_2}{A_2}\, = \,
  -\frac{3(\ell+2)(2\ell^2+5\ell+4)}{(2\ell+3)(2\ell+1)},
  \label{ratio-B2A2-extremal}
\end{equation}
using Eqs.~\eqref{Y2-extremal-Lge2} and \eqref{Ypt-extremal-Lge2}.
Since the formulas of $B_2/A_2$
for sub- and super-extremal cases are complicated,
we show their values in the cases $\ell=2$ and $3$
as functions of $Q/M$ by the graph of 
Fig.~\ref{Fig:DrawRatioLge2_combined}.
For each value of $\ell$, there are two branches
that corresponds to the values of $B_2/A_2$ evaluated
at the outer and inner photon surfaces, $r=r_{\rm p}^{(\pm)}$,
and the two branches merge at $Q/M=\sqrt{9/8}$.

Here, we evaluate interesting quantities on the photon surface,
$H^{(1)}(r_{\rm p})$, $K^{(1)}(r_{\rm p})$, and $\varphi^{(1)}(r_{\rm p})$.
If $H^{(1)}(r_{\rm p})\neq 0$ and $\varphi^{(1)}(r_{\rm p})\neq 0$, the lapse function
and the electrostatic potential 
are not constant on $r=r_{\rm p}$, respectively.
From the uniqueness theorems for the (redefined) photon sphere of
Refs.~\cite{Cederbaum:2015b,Yazadjiev:2015a},
at least one of these two quantities must be nonzero
on the distorted photon surface.
The quantity $K^{(1)}(r_{\rm p})$  is related
to the intrinsic geometry
of the two-dimensional section of the $t=\mathrm{constant}$ slice and
the photon surface, because its  Ricci scalar ${}^{(2)}R$ is given by
\begin{equation}
  {}^{(2)}R \ = \ \frac{2}{r_{\rm p}^2}
  \left[1+\epsilon(\ell^2+\ell-2)K^{(1)}(r_{\rm p})Y_{\ell}^0(\theta)\right],
\end{equation}
for the axisymmetric mode. 
Therefore, as long as $K^{(1)}\neq 0$ on the photon surface,
the spatial section of the photon surface is distorted.
Each of the solutions for $H^{(1)}$, $K^{(1)}$, and $\varphi^{(1)}$
has the parameters $A_2$ and $B_2$,
and we eliminate $B_2$ using the relation of Eq.~\eqref{PS-condition-Lge2}.
Then, all these quantities are proportional to $A_2$. 
Figure~\ref{Fig:Ricci2D_Lge2_combined} shows the values
of ${H^{(1)}}/{A_2}$ (top left), ${K^{(1)}}/{A_2}$ (top right), and
${\varphi^{(1)}}/{A_2}$ (bottom) 
on the photon surface as functions of $Q/M$
for $\ell=2$ and $3$. The values of these quantities are always negative.
This means that our results are consistent with
the uniqueness theorem for the photon sphere,
and that the photon surface is actually
distorted.

To summarize, there are two independent regular solutions
to each of the $\ell\ge 2$ modes, and if we tune their amplitude
so that Eq.~\eqref{PS-condition-Lge2} is satisfied, the surface
$r=r_{\rm p}$ satisfies the photon surface condition.
This means that the distorted photon surface can be
formed in electrovacuum spacetimes. The lapse function
and the electrostatic potential are not constant on it.

\subsection{The case $\ell=1$}

\begin{figure}
  \centering
  \includegraphics[width=0.5\textwidth,bb=0 0 294 269]{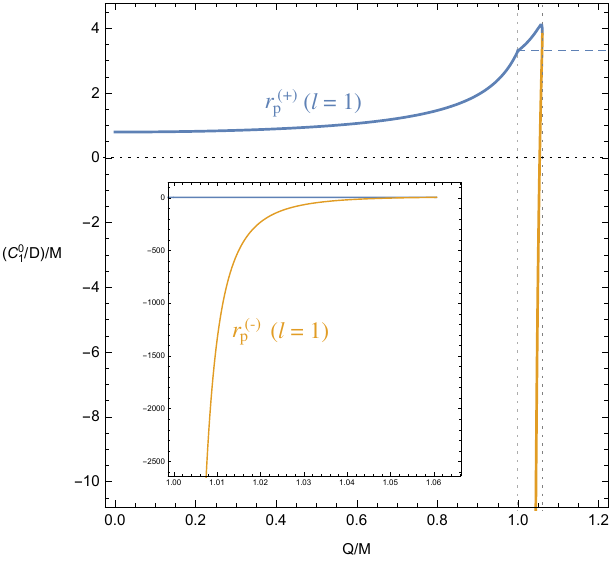}
  \caption{The values of $(C_1^0/D)/M$ that realize the 
    photon surface at $r_{\rm p}^{(+)}$ and $r_{\rm p}^{(-)}$ for $\ell=1$ 
    as functions of $Q/M$. There are two branches, and
    the upper and lower branches correspond to the outer and
    inner photon surfaces, respectively. The 
    horizontal dashed line indicates $(C_1^0/D)/M=33/10$
    in the extremal case. The inset
  shows the large-scale behavior of the lower branch.}
  \label{Fig:DrawRatioLeq1_combined}
\end{figure}

We now turn our attention to the case of $\ell = 1$.
There are the modes $(\ell, m) = (1,1)$, $(1,0)$, and $(1,-1)$,
and below, the axisymmetric mode $(\ell, m)=(1,0)$ is considered
since the other two modes can be obtained by rotating the axisymmetric
mode. In the perturbative solution
for the mode $\ell=1$, there is only one physical solution 
as discussed in Sec.~\ref{Sec:Solving_equations_L1},
and this is in contrast to the cases $\ell \ge 2$
where there are two physical modes that enable the formation
of a distorted photon surface.
However, in contrast to the cases $\ell\ge 2$,
there is one degree of freedom of shifting the
coordinate position of the photon surface
as $f^{(1)}=C_1^0 Y_1^0(\theta)$.
Since $f^{(1)}_{,\theta\theta}=-f^{(1)}$ holds,
Eq.~\eqref{ps-eq3} implies that
the photon surface condition is satisfied
if we choose $C_1^0$ and $D$ such that
\begin{equation}
  \frac{C_1^0}{D} \ = \ \left.-\frac{r^2}{6}\left(h^{(1)}_{\mathrm{phys}}+k^{(1)}_{\mathrm{phys}}\right)_{,r}\right|_{r=r_{\rm p}}
  \label{PhotonSurfaceCondition-L1}
\end{equation}
is satisfied, where we used the fact that 
$\nu^{(0)}_{,rr}+r^{-2}+{r^{-2}e^{-2\nu^{(0)}}}=-6/r^2$
holds on the background photon surface $r=r_{\rm p}$. 
In the extremal case, this condition is rewritten as
$C_0^1/D = (33/10)M$ using Eqs.~\eqref{Physical-part-extremal-H1-L1}
and \eqref{Physical-part-extremal-K1-L1}.
Since the formulas for sub- and super-extremal cases
are complicated, we show the behavior of $(C_1^0/D)/M$
as functions of $Q/M$ in Fig.~\ref{Fig:DrawRatioLeq1_combined}.
Again, there are two branches
that corresponds to the values of $(C_1^0/D)/M$ evaluated
at the outer and inner photon surfaces, $r=r_{\rm p}^{(\pm)}$,
and the two branches merge at $Q/M=\sqrt{9/8}$.
For a positive $D$, the value of $C_1^0$ is 
positive in the branch of the outer photon surface,
while $C_1^0$ changes its sign in the branch of the
inner photon surface. We will discuss the reason
at the last of this section.

%
%

Let us discuss the properties of the
photon surface. Taking account of the shift in the coordinate position,
the induced metric of the photon surface
can be written as
\begin{equation}
  ds^2 \ =\ -e^{2\nu^{(0)}}(r_{\rm p})
  \left[1-2\epsilon\widehat{H}^{(1)}Y_1^0(\theta)\right]dt^2
  +r_{\rm p}^2
  \left[1+2\epsilon\widehat{K}^{(1)}Y_1^0(\theta)\right](d\theta^2+\sin^2\theta d\phi^2),
\end{equation}
where
\begin{subequations}
\begin{eqnarray}
\widehat{H}^{(1)} &=& H^{(1)}(r_{\rm p})-\frac{C_1^0}{r_{\rm p}},\\
\widehat{K}^{(1)} &=& K^{(1)}(r_{\rm p})+\frac{C_1^0}{r_{\rm p}}.
\end{eqnarray}
\end{subequations}
We first consider the section of a $t=\mathrm{constant}$ surface
and the photon surface. 
The value of the two-dimensional Ricci scalar ${}^{(2)}R$
is not changed to first order, and hence, the geometry of the spatial section
of the photon sphere
remains spherically symmetric and is not distorted.
Next, let us consider the lapse function. 
Using Eq.~\eqref{formal-solution-H1-L1} with $b=0$ and
Eq.~\eqref{PhotonSurfaceCondition-L1}, $\widehat{H}^{(1)}$ is
proportional to $D$ as
\begin{equation}
  \widehat{H}^{(1)} \ = \ D\left.\left[h^{(1)}_{\rm phys}+\frac{r_{\rm p}}{6}
    \left(h^{(1)}_{\mathrm{phys}}+k^{(1)}_{\mathrm{phys}}\right)_{,r}\right]\right|_{r=r_{\rm p}}.
\end{equation}
If $\hat{H}^{(1)}$ is not zero, the lapse function is not constant
on the photon surface.
Another interesting quantity is the
electrostatic potential on the photon surface.
It is written as
\begin{equation}
\Phi \ = \ \frac{Q}{r_{\rm p}}+\widehat{\varphi}^{(1)} Y_1^0(\theta),
\end{equation}
where
\begin{equation}
  \widehat{\varphi}^{(1)}\ = \ {\varphi}^{(1)}(r_{\rm p})
  -\frac{Q}{r_{\rm p}^2}C_1^0.
  \label{widehat_varphi_intermediate}
\end{equation}
Equations~\eqref{varphi_L1_extremal}, \eqref{varphi_L1_subextremal},
and \eqref{varphi_L1_superextremal} are
symbolically written as ${\varphi}^{(1)} = D{\varphi}_D^{(1)}$ in the case
$\beta=0$, and substituting this equation and 
Eq.~\eqref{PhotonSurfaceCondition-L1} into
Eq.~\eqref{widehat_varphi_intermediate},
$\widehat{\varphi}^{(1)}$ is also proportional to $D$ as
\begin{equation}
  \widehat{\varphi}^{(1)}\ = \ \left.
  D\left[{\varphi}_D^{(1)}+\frac{Q}{6}\left(h^{(1)}_{\mathrm{phys}}+k^{(1)}_{\mathrm{phys}}\right)_{,r}\right]\right|_{r=r_{\rm p}}.
\end{equation}

%
\begin{figure}
  \centering
  \includegraphics[width=0.4\textwidth,bb=0 0 294 285]{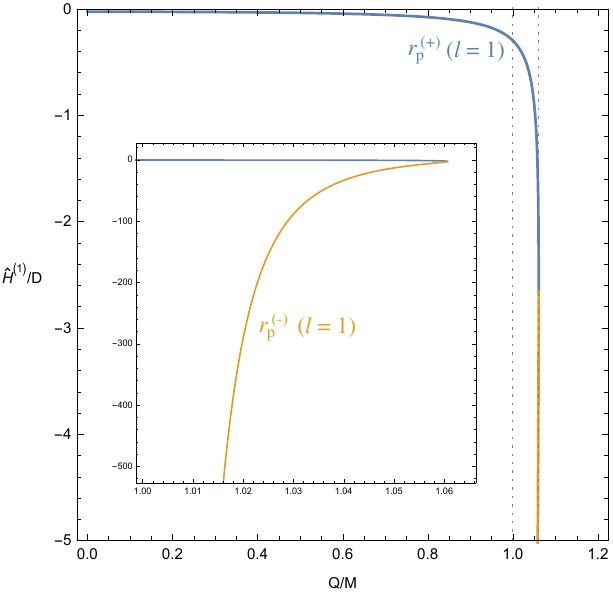}
  \includegraphics[width=0.4\textwidth,bb=0 0 294 281]{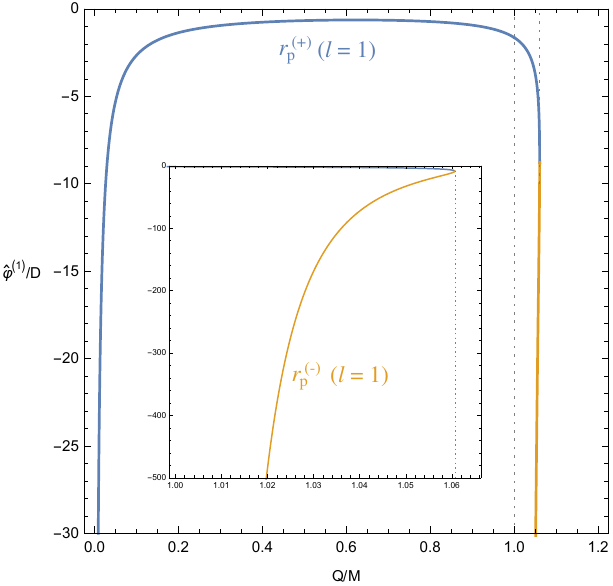}
  \caption{The values of $\widehat{H}^{(1)}/D$ (left)
    and $\widehat{\varphi}^{(1)}/D$ (right)
    on the 
    photon surfaces for $\ell=1$ 
    as functions of $Q/M$. In each panel, there are two branches, and
    the lower and upper branches correspond to the outer and
    inner photon surfaces, respectively. 
    Each inset shows the global behavior of the lower branch.
    $\widehat{H}^{(1)}/D$ and 
    $\widehat{\varphi}^{(1)}/D$ are always negative.}
  \label{Fig:Lapse_Leq1_combined}
\end{figure}
%

Figure~\ref{Fig:Lapse_Leq1_combined} shows the value of
$\widehat{H}^{(1)}/D$ (left panel) and $\widehat{\varphi}^{(1)}/D$
(right panel) as functions of $Q/M$. 
In each panel, there are two branches that corresponds to
the outer and inner photon surface.
For all cases, the value of $H^{(1)}$ is negative,
and hence, the lapse function is not constant on the photon surface.
This means that the norm of the timelike Killing field
is not constant. Due to this property, the photon surface
is distorted in the three-dimensional sense,
although the two-dimensional spatial section is spherically symmetric.
Similarly, the value of $\widehat{\varphi}^{(1)}(Q/M)$
is always negative. 
In any case, since the lapse function and the electrostatic potential
do not become zero, our results are
consistent with the uniqueness theorems for the photon sphere
of \cite{Yazadjiev:2015a,Cederbaum:2015b}.

This result may be interpreted as follows.
On the one hand, 
let us consider the outer photon surface at $r=r_{\rm p}^{(+)}$,
in the case that $Q$ and $D$ are positive.
In this case, the electric dipole moment
which is directed downward (i.e. the direction of $\theta=\pi$)
is perturbatively added.
This means that the electric field becomes stronger and weaker
at $\theta=\pi$ and $0$, respectively.
Let us recall the fact that if the electric field
becomes stronger outside of some region,
the Komar energy of that region becomes smaller,
and hence, gravity becomes weaker.
This would indicate that gravity around $\theta=\pi$ becomes weaker,
while that around $\theta=0$ becomes stronger.
This is consistent with the fact that $H^{(1)}$ is positive,
because in the presence of timelike Killing field, 
$-(g_{tt}+1)\approx e^{2\nu^{(0)}}[1-2\epsilon H^{(1)}(r)Y(\theta)]-1$ has an analogy to the Newton potential.
This makes the photon surface shift upward (i.e., the direction of $\theta=0$),
and thus, $C_1^0$ becomes positive
in the branch of the outer photon surface
in Fig.~\ref{Fig:DrawRatioLeq1_combined}.
On the other hand, around the inner photon surface at $r=r_{\rm p}^{(-)}$,
background gravity becomes weaker as the radius $r$ is decreased.
This would cause the flip
of the sign of $C_1^0$ in the branch of the inner photon surface
in Fig.~\ref{Fig:DrawRatioLeq1_combined}.

To summarize, it is possible to have a photon surface
in a perturbed Reissner-Nordstr\"om spacetime
for the $\ell=1$ mode as well. 
Although the spatial section of the photon surface is spherically symmetric,
the photon surface is distorted in the three-dimensional sense
since the lapse function is not constant.
The electrostatic potential also changes its value on the distorted photon surface.

%
%
\section{Summary and discussion}
\label{Sec:conclusion}

In this paper, we have studied whether a static photon surface
can be present in a perturbed Reissner-Nordstr\"om spacetime.
The analytic solutions to the equations for the static perturbations
are solved assuming the regularity
around and outside the background photon surface.
For each mode of $\ell\ge 2$, there are
two independent solutions that correspond to adding
the multipole moments of mass and electric charge, respectively.
By adjusting the two amplitudes of the two independent solutions,
the existence of a distorted photon surface is realized.
Although there is only one physical solution for the mode $\ell = 1$,
there is also a degree of freedom of shifting the coordinate position
of the photon surface, and a photon surface can be present by
adjusting its position. 
Although the spatial section of that photon surface
is spherically symmetric, it is distorted
in the three-dimensional sense because the lapse is not constant.
Our results indicate that the distorted photon surface can
be realized at least at the level of the first-order perturbation
of the Reissner-Nordstr\"om spacetime.

The results of this paper may seem to be in contrast
to our previous paper \cite{Yoshino:2016} where the present author
has proved the perturbative uniqueness of the photon surface
in vacuum spacetimes. 
We would like to note that in that paper, the possibility of the
photon surface formation is pointed out
in the situation where matter is present in the outside region,
although a fine-tuned distribution of matter is
required. In the system of this paper, the electric fields are present
instead of matter, and by adjusting the perturbation
generated by the electric fields, the
formation of a distorted photon surface has become possible.
Therefore, the conclusion of this paper is not contradictory
to our previous result of \cite{Yoshino:2016}.

Note that the results of this paper 
do not guarantee the existence of a photon surface
in the situation where full nonlinearity of the Einstein equation
is taken into account. Here, let us briefly discuss
whether the second-order perturbation prevents the existence
of the photon surface or not.
The higher-order perturbation of the Reissner-Nordstr\"om
spacetime is formulated in Ref.~\cite{Rutkowski:2019}.
For simplicity, let us consider the first-order perturbation 
in the $\ell=2$ mode. At the second-order perturbation,
the products of the first-order perturbation quantities
generate the second-order perturbation
in the $\ell=0$, $2$, and $4$ modes.
As for the change caused by the $\ell=0$ mode,
it would be possible to
satisfy the photon surface condition 
by adjusting the radial position
of the photon surface.  
As for the change caused by the $\ell=2$ and $4$ modes,
there are degrees of freedom of adding the homogeneous
solutions to the second-order Einstein-Maxwell equations,
which have the same solutions as the first-order perturbation.
Then, the photon surface could be satisfied by adjusting
the amplitudes of the homogeneous solutions. 
In this way, it seems that there would be no difficulty
in making a photon surface after the higher-order perturbations
are taken into account.
For this reason, we conjecture that
a photon surface can be present in fully nonlinear electrovacuum spacetimes.
In other words, we suspect that the uniqueness
may not hold for the photon surface in electrovacuum spacetimes,
although we do expect that the uniqueness would hold in vacuum spacetimes
as conjectured in our previous paper \cite{Yoshino:2016}.
Of course, the direct examination
by analytic or (highly accurate) numerical calculations
are required to derive a rigorous conclusion.

Since the distorted static photon surface
seems to be formed in principle,
it would be interesting to consider
the implication for the phenomena in the astrophysical context.
If the self gravity of a black hole accretion disk
is considered, there would be situations with and without
the photon surface because there are many models of the accretion disks.
It was numerically clarified
in Ref.~\cite{Shoom:2017} that if the photon surface is absent,
the behavior of null geodesics becomes complex, and the photon
sphere transforms into a ``fractal basin boundary''.
If the photon surface is present, such complex 
behavior would not happen because the photon surface is umbilical,
and hence, the geometry around
the photon surface is relatively simple.
For this reason, phenomena that are unique to the presence of
the photon surface may happen, and
it might be possible to extract the information
of the accretion disks by observing such phenomena. 
Exploring such a possibility would be an interesting remaining issue.

%
%
\acknowledgments

The author thanks Ken-ichi Nakao and Chul-moon Yoo for helpful comments.
H. Y. is in part supported by JSPS KAKENHI Grant Numbers JP22H01220 and
JP21H05189,
and is partly supported by Osaka Central Advanced Mathematical Institute 
(MEXT Joint Usage/Research Center on Mathematics and Theoretical Physics JPMXP0723833165).

\appendix
%
%
\section{Explicit formulas for the associated Legendre functions}
\label{Appendix_A}

In this Appendix, we present the explicit formulas
for the associated Legendre functions of the first and second kinds
used in the perturbative analysis of sub- and super-extremal cases
in Secs.~\ref{Sec:Solving_the_perturbation_equations} and \ref{Sec:Solving_equations_L1}.

\subsection{$P_\ell^1(z)$ and $Q_{\ell}^1(z)$}
\label{App:Explicit-P-Q}

In the sub-extremal cases, we have presented the formulas for
${P}_\ell^1(z)$ and ${Q}_{\ell}^1(z)$
in Eqs.~\eqref{LegendreP-standard} and \eqref{LegendreQ-standard}.
Their explicit formulas are
\begin{subequations}
\begin{eqnarray}
{P}_1^1(z)&=&\sqrt{z^2-1},\\
{P}_2^1(z)&=&3z\sqrt{z^2-1},\\
{P}_3^1(z)&=&\frac32(5z^2-1)\sqrt{z^2-1},
\end{eqnarray}
\end{subequations}
and
\begin{subequations}
\begin{eqnarray}
{Q}_1^1(z)&=&-\frac{z}{\sqrt{z^2-1}}+\sqrt{z^2-1}\, \mathrm{arccoth}(z),\\
{Q}_2^1(z)&=&-\frac{3z^2-2}{\sqrt{z^2-1}}+3z\sqrt{z^2-1}\, \mathrm{arccoth}(z),\\
{Q}_3^1(z)&=&-\frac{z(15z^2-13)}{2\sqrt{z^2-1}}
+\frac32 (5z^2-1)\sqrt{z^2-1}\, \mathrm{arccoth}(z),
\end{eqnarray}
\end{subequations}
for $\ell=1$, $2$, and $3$, respectively.
Note that
\begin{equation}
\mathrm{arccoth}(z) \ = \ -\frac12\log\left(\frac{z-1}{z+1}\right)
\end{equation}
holds for the range $z>1$, which corresponds to the region outside
the event horizon in the setup of \ref{Sec:SubExtremal-Lge2}
and \ref{Sec:SubExtremal-Leq1}.

\subsection{$\tilde{P}_\ell^1(z)$ and $\tilde{Q}_{\ell}^1(z)$}
\label{App:Explicit-tilP-tilQ}

In the super-extremal cases, we have defined
the functions $\tilde{P}_\ell^1(z)$ and $\tilde{Q}_{\ell}^1(z)$
in Eqs.~\eqref{Def:TildeP} and \eqref{Def:TildeQ}.
Their explicit formulas are
\begin{subequations}
\begin{eqnarray}
\tilde{P}_1^1(z)&=&\sqrt{z^2+1},\\
\tilde{P}_2^1(z)&=&3z\sqrt{z^2+1},\\
\tilde{P}_3^1(z)&=&\frac32(5z^2+1)\sqrt{z^2+1},
\end{eqnarray}
\end{subequations}
and
\begin{subequations}
\begin{eqnarray}
\tilde{Q}_1^1(z)&=&-\frac{z}{\sqrt{z^2+1}}+\sqrt{z^2+1}\, \mathrm{arccot}(z),\\
\tilde{Q}_2^1(z)&=&\frac{3z^2+2}{\sqrt{z^2+1}}-3z\sqrt{z^2+1}\, \mathrm{arccot}(z),\\
\tilde{Q}_3^1(z)&=&-\frac{z(15z^2+13)}{2\sqrt{z^2+1}}
+\frac32 (5z^2+1)\sqrt{z^2+1}\, \mathrm{arccot}(z),
\end{eqnarray}
\end{subequations}
for $\ell=1$, $2$, and $3$, respectively.

%
%
\section{Explicit formulas for $Y_{\rm pt}$}
\label{Appendix_B}

In Sec.~\ref{Sec:Solving_the_perturbation_equations},
we have presented the formal solutions for $Y_{\rm pt}$
in Eqs.~\eqref{Ypt-subextremal-Lge2} and
\eqref{Ypt-superextremal-Lge2} in the sub- and super-extremal
cases, respectively. 
Here, we present the specific formulas for $Y_{\rm pt}$
after ingtegration for $\ell=2$ and $3$.
As quantities that are necessary for the calculation of $Y_{\rm pt}$,
we note
\begin{subequations}
\begin{eqnarray}
q_{21} \ = \ -\tilde{q}_{21} & = & -\frac25,\\  
q_{31} \ = \ -\tilde{q}_{31} & = & -\frac{8}{35},
\end{eqnarray}
\end{subequations}
where $q_{\ell\mu}$ and $\tilde{q}_{\ell\mu}$ is defined in Eqs.~\eqref{q_ellmu}
and \eqref{tilq_ellmu}.

\subsection{The sub-extremal case}
\label{App:SubExtremal-Ypt}

For the sub-extremal case, the formula for $Y_{\rm pt}$
that is obtained after integration of Eq.~\eqref{Ypt-subextremal-Lge2} is
\begin{equation}
  Y^{(\ell=2)}_{\rm pt} \ = \
  -\frac{5}{2\gamma^4}\left[6 - \frac{1}{z^2-1}
  +\frac{15z^2-13}{\gamma z+1}
  -\frac{3(5z^3+2\gamma z^2 -4z -\gamma)}{\gamma z + 1}
  \,\mathrm{arccoth}(z)\right],
\end{equation}
for $\ell=2$, and is
\begin{multline}
  Y^{(\ell=3)}_{\rm pt} \ = \ -\frac{35}{16\gamma^5(\gamma z + 1)}
  \bigg[
  \frac{105z^5+45\gamma z^4-175z^3-63\gamma z^2+68 z +16\gamma}{z^2-1}
  \\
  -3(35z^4+15\gamma z^3 -35z^2-11\gamma z +4)
  \,\mathrm{arccoth}(z)\bigg],
\end{multline}
for $\ell=3$.

\subsection{The super-extremal case}
\label{App:SuperExtremal-Ypt}

For the super-extremal case, the formula for $Y_{\rm pt}$
that is obtained after integration of Eq.~\eqref{Ypt-superextremal-Lge2} is
\begin{equation}
  Y^{(\ell=2)}_{\rm pt} \ = \
  \frac{5}{2\gamma^4}\left[-6 - \frac{1}{z^2+1}
  +\frac{15z^2+13}{\gamma z+1}
  -\frac{3(5z^3-2\gamma z^2 +4z -\gamma)}{\gamma z + 1}
  \,\mathrm{arccot}(z)\right],
\end{equation}
for $\ell=2$, and is
\begin{multline}
  Y^{(\ell=3)}_{\rm pt} \ = \
  \frac{35}{16\gamma^5(\gamma z + 1)}
  \bigg[
  \frac{-105z^5+45\gamma z^4-175z^3+63\gamma z^2-68 z +16\gamma}{z^2+1}
  \\
  +3(35z^4-15\gamma z^3 +35z^2-11\gamma z +4)
  \,\mathrm{arccot}(z)\bigg],
\end{multline}
for $\ell=3$.

%
%
\section{Explicit formulas for $\varphi^{(1)}$}
\label{Appendix_C}

In this appendix, 
we present the specific formulas for the electrostatic potential
$\varphi^{(1)}$ for $\ell=2$ and $3$.
For simplicity, we set
\begin{equation}
\varphi^{(1)} \ = \ A_2\varphi^{(1)}_A + B_2\varphi^{(1)}_B.
\end{equation}

\subsection{The sub-extremal case}
\label{App:SubExtremal-varphi}

For the sub-extremal case, the formulas for $\varphi_{A}$ and
$\varphi_{B}$ are
\begin{subequations}
\begin{multline}
  \varphi^{(1)}_A \ = \
-\frac{5}{4\gamma^3\sqrt{1-\gamma^2}(\gamma z+1)^2}\left\{
  3(5-\gamma^2)z^3+12\gamma z^2-(13-5\gamma^2)z-8\gamma\right.
  \\ \left.
  -3(z^2-1)[(5-\gamma^2)z^2+4\gamma z-1+\gamma^2]\mathrm{arccoth}(z)\right\},
\end{multline}
\begin{equation}
  \varphi^{(1)}_B \ = \ -\frac{5(z+\gamma)[-3z^2+2+3z(z^2-1)\mathrm{arccoth}(z)]}{4\gamma\sqrt{1-\gamma^2}(\gamma z+1)^2},
\end{equation}
\end{subequations}
for $\ell=2$, and are
\begin{subequations}
\begin{multline}
  \varphi^{(1)}_A \ = \ -\frac{35}{32\gamma^4\sqrt{1-\gamma^2}(\gamma z+1)^2}\left\{
  15(7-2\gamma^2)z^4+75\gamma z^3-5(23-10\gamma^2)z^2-65\gamma z\right.
  \\ \left.
  +16(1-\gamma^2)
  -15(z^2-1)[(7-2\gamma^2)z^3+5\gamma z^2-(3-2\gamma^2) z-\gamma]\mathrm{arccoth}(z)\right\},
\end{multline}
\begin{equation}
  \varphi^{(1)}_B \ = \ -\frac{35(z+\gamma) [-z(15z^2-13)+3(5z^4-6z^2+1)\mathrm{arccoth}(z)]}{32\gamma^2\sqrt{1-\gamma^2}(\gamma z+1)^2},
\end{equation}
\end{subequations}
for $\ell=3$.

\subsection{The super-extremal case}
\label{App:SuperExtremal-varphi}

For the super-extremal case, the formulas for $\varphi_{A}$ and
$\varphi_{B}$ are
\begin{subequations}
\begin{multline}
  \varphi^{(1)}_A \ = \ \frac{5}{4\gamma^3\sqrt{1+\gamma^2}(\gamma z+1)^2}\left\{
  3(5+\gamma^2)z^3-12\gamma z^2+(13+5\gamma^2)z-8\gamma
  \right.
  \\ \left.
  -3(z^2+1)[(5+\gamma^2)z^2-4\gamma z+1+\gamma^2]\mathrm{arccot}(z)\right\},
\end{multline}
\begin{equation}
  \varphi^{(1)}_B \ = \ \frac{5(z-\gamma)[3z^2+2-3z(z^2+1)\mathrm{arccot}(z)]}{4\gamma\sqrt{1+\gamma^2}(\gamma z+1)^2},
\end{equation}
\end{subequations}
for $\ell=2$, and are
\begin{subequations}
\begin{multline}
  \varphi^{(1)}_A \ = \ \frac{35}{32\gamma^4\sqrt{1+\gamma^2}(\gamma z+1)^2}\left\{
  -15(7+2\gamma^2)z^4+75\gamma z^3-5(23+10\gamma^2)z^2+65\gamma z
  \right.
  \\ \left.
  -16(1+\gamma^2)
  +15(z^2+1)[(7+2\gamma^2)z^3-5\gamma z^2+(3+2\gamma^2) z-\gamma]\mathrm{arccot}(z)\right\},
\end{multline}
\begin{equation}
  \varphi^{(1)}_B \ = \ \frac{35(z-\gamma) [-z(15z^2+13)+3(5z^4+6z^2+1)\mathrm{arccot}(z)]}{32\gamma^2\sqrt{1+\gamma^2}(\gamma z+1)^2},
\end{equation}
\end{subequations}
for $\ell=3$.



\end{document}